\documentclass[12pt,headings=small,usenatbib]{article}
\pdfoutput=1
\usepackage{newtxtext,newtxmath}
\usepackage[T1]{fontenc}
\usepackage{ae,aecompl}
\DeclareRobustCommand{\VAN}[3]{#2}
\let\VANthebibliography\thebibliography
\def\thebibliography{\DeclareRobustCommand{\VAN}[3]{##3}\VANthebibliography}
\usepackage{graphicx}
\usepackage{amsmath}
\usepackage[authoryear]{natbib}
\usepackage{url}
\usepackage{xcolor}
\usepackage{pdflscape}
\newcommand{\beq}{\begin{equation}}
\newcommand{\eeq}{\end{equation}}
\newcommand{\beqa}{\begin{eqnarray}}
\newcommand{\eeqa}{\end{eqnarray}}
\def\lap{\lower.5ex\hbox{$\; \buildrel < \over \sim \;$}}
\def\gap{\lower.5ex\hbox{$\; \buildrel > \over \sim \;$}}

\begin{document}

\centerline{\Large{The Physicists' Philosophy of Physics}}\medskip

\centerline{P. J. E. Peebles\footnote{E-mail: pjep@Princeton.edu}}
\centerline{Joseph Henry Laboratories}
\centerline{Princeton University, Princeton, NJ, USA}

\begin{abstract}
\noindent Research in physics operates by an implicit community philosophy.  I offer a definition I think physicists would accept, by and large, and explanations by examples from physics and comparisons to what philosophers, sociologists, and historians as well as physicists conclude from what they observe physicists doing.
\end{abstract}

\section{Introduction}

Physicists follow operating procedures that have been close enough to standard for long enough to constitute a working philosophy. I do not have in mind the methods of discovery of specific physical theories and the tests that establish some as canonical physics, both of which which have evolved a lot, but rather our canonical practice. Experience can force evolution of what is canonical, but it has been stable long enough to merit a careful definition of the physicists' way of thinking: the physicists' homegrown philosophy of physics.

 Natural scientists seek empirically tested regularities that characterize what is observed. In fundamental physics, the subject of this essay, the regularities are described by mathematical theories that are meant to be usefully close to what is assumed to be the foundational basis for how the physical world operates. That it can succeed is an assumption that is implicit in what physicists do and encouraged by experience, but seldom stated. I offer in Section~\ref{sec:startingassumptions}  a statement of this assumption and what I take to be the other three basic elements of the physicists' philosophy of physics. These assumptions need interpretations; it is the subject of Section~\ref{sec:interpretations}. Research in physics has arrived at good approximations to what is assumed to be the fundamental nature of mind-independent reality. Section~\ref{sec:tests} is a review evidence of this from the comparisons of what our theories predict and what is measured or observed. It fits the assumptions in Section~\ref{sec:startingassumptions} so far. Physics has a sociology, though physicists seldom think about it. I offer thoughts about this along with assessments of what sociologists say about physics in Section~\ref{sec:sociology}. The mixed opinions about the future of research in fundamental physics are reviewed in Section~\ref{sec:final theory}. Section~\ref{Concluding thoughts} presents concluding remarks.

Scattered notes: The topic of this essay is the philosophy of research aimed at advancing the physics that is the closest approximation we have to physical reality, assuming it exists. The other physical sciences that have theories capable of predictions that can be compared to experiments, and the natural sciences that deal with observations, have similar philosophies though of course adapted to be suited to the working conditions. These categories of science cannot be ranked by intrinsic merit, but fundamental physics has gone particularly far because it deals with particularly simple analyzable situations. I do not mean to trespass on thinking about natural science by philosophers and sociologists, or add to the literature on ontology by my comments about the physicists' (usually only implicit) concept of reality. My purpose is to present what I take to be the framework of ideas implicit in the culture of research into fundamental physical reality. The four starting assumptions to be stated next (in Sec.~\ref{sec:startingassumptions}) are meant to be succinct statements of this philosophy. They might be of interest to philosophers and sociologists who judge what people are doing in science, or maybe not. But I think is useful to have a careful statement of the  philosophy of physicists as we find ourselves practicing it.

\section{Starting Assumptions}\label{sec:startingassumptions}

The starting idea of the natural sciences is that the world operates by rules and regularities that can be discovered by measurements: experiments, observations, descriptions. In fundamental physics the idea can be narrowed to four starting assumptions.\vspace{1.5ex}

A: The world operates by rules and the logic of their application that can be discovered in successive approximations.

B: A useful approximation to the rules, a theory, yields predictions that are reliably computed or described and found to agree with reliable measurements or observations, within the uncertainties of the predictions and measurements.

C: Fundamental physics is growing more complete by advances in the variety and precision of independent empirical measurements that check and agree with predictions, and by occasional unifications that demote well-tested fundamental physical theories to good approximations to still better more broadly tested theories.

D: Research in fundamental physical science is advancing toward a unified mind-independent physical reality, though not  one that need ever be completely known.\vspace{1.5ex}

In correspondence received about the draft of this essay on arXiv and elsewhere, these statements have been termed ``axioms,'' or ``beliefs,'' or ``faiths.'' All make sense, but to me the connotation of confidence feels too strong. I prefer ``assumptions,'' or ``hypotheses,'' because the statements are subject to adjustment if the empirical evidence requires it. I use the word ``know'' only in the informal sense, that ``it's pretty clear.'' Correspondents also have wondered why the assumptions make no mention of how specific theories were discovered and came to be accepted or rejected, and why there is scant attention to this in the following discussion. It is because my topic is the framework, or guidelines, for curiosity-driven scientific research implicitly defined by what we find ourselves doing that works, with particular attention to what has happened in fundamental physics on very small and very large scales. 

\section{Interpreting the Assumptions}\label{sec:interpretations}

The four assumptions in Section~\ref{sec:startingassumptions} require interpretations: judgements of intent, policies for ambiguities, and allowances for sensible exceptions, all informed by experience. I draw lessons about this from the history of physics since Maxwell unified the theories of electricity and magnetism. More is to be learned from the history of how natural science grew up to the time of Newton, but only occasional thoughts about earlier advances are offered here. 

\subsection{Rules and logic}\label{sec:ruleslogic}

The starting thought in Assumption~A in Section~\ref{sec:startingassumptions} is that the world operates by rules we can discover. This bold idea is implicit in scientists' expectation that their research is adding to what is known about the world. We have no guarantee that this is so, of course, but it is encouraged by experience. 

The topic of research might be birds. There are distinct observational differences among the species of birds, with the interesting exceptions of crossbreeds, and I am told that genetic testing by and large confirms these visual classifications, with rare exceptions that are big news to bird watchers. It reenforces the ornithologists' implicit assumption that they are discovering aspects of the rules by which the world operates. The Ptolemaic system produces remarkably accurate predictions of the angular positions of the planets. This also amounts to useful rules for how the world operates, but the system says nothing about how an apple falls to the ground. The Newtonian theory takes care of falling apples, the precession of Foucault's pendulum, the flow of water in tides and air in tornados, and so much more: a far more unified approximation to reality. The equations Maxwell wrote down in the mid-1800s are tested and prove to be sound by everyday experience. That is not what you would expect in a universe that does not operate by fixed rules. For that matter, how could life form and how could the fittest survive if Nature kept changing what is required to be fit? All the many considerations of this sort are commonplace evidence that Nature is operating by rules we can discover and rely on. 

The rules expressed as mathematical laws in fundamental physics, the theories, are supposed to approximate the fundamental basis for how our world operates. The demonstration that this is so cannot be compared to a mathematician's proof of a theorem. It is instead the accumulation of checks of theory by observation, some precise, some qualitative but  important, in situations that are simple enough to allow tests. In a situation that is too complicated to apply tests we don't know for sure the  theory is a good approximation; it is a loose end that awaits further work. I comment on this in Section~\ref{sec:final theory}.

The mathematical nature of the rules was not always so important; Faraday made great discoveries about the properties of electricity and magnetism without the use of mathematics much beyond arithmetic. But then Maxwell added the hypothetical displacement current to arrive at a mathematical theory that unified two subjects, electricity and magnetism, to get the theory of electromagnetism that has given us so many practical applications. But we cannot know the future of fundamental physics; maybe it will involve something beyond the mathematics now known; so it seems best to use the more generic term ``rule'' rather than ``theory'' in Assumption~A.

Quantum physics taught us a cautionary lesson about the logic mentioned in Assumption~A. An electron acts like a particle in many ways, but in standard quantum physics an electron can be placed in a pure state of two spatially separated wave packets, two entangled states, as in a double slit experiment. It seems logical to insist that the electron, being a particle, really is in just one of the two wave packets. The physicist John Stewart Bell argued for this logic, and for versions of David Bohm's theory of an invisible pilot wave that allows the electron to act as if it were in a quantum entangled state of the two wave packets yet really be present in only one of the packets. Normal logic demands it. Bell gave us a celebrated test to distinguish this from the logic of quantum physics. In the essay, {\it Bertlmann's socks and the nature of reality}, Bell (1981) explained how to test between pilot waves and the standard quantum physics. Bell mentioned that the physicist Alain Aspect ``is engaged in an experiment'' applying the test. John Francis Clauser had led earlier experimental tests. The results of these and many other experiments consistently reject pilot waves and agree with quantum physics. We must adjust our logic of how the world operates, for now at least.

Bell (1976) was troubled by the absence of a 
\begin{quotation}
\noindent sharply defined boundary between what is to be treated as microscopic and what as macroscopic, and this introduces a basic vagueness into fundamental physical theory. But this vagueness, because of the immense difference of scale between the atomic level where quantum concepts are essential and the macroscopic level where classical concepts are adequate, is quantitatively insignificant in any situation hitherto envisaged. So, it is quite acceptable to many people.
\end{quotation}
Bell's distinction between microscopic and macroscopic is important but not always that pronounced. The electrons in the coherent state of a superconductor with macroscopic dimensions, a meter around, are entangled in Cooper pairs that do not have definite positions. The standard arguments for the reconciliation of this with classical physics are reasonable but not totally convincing. We must fall back on Assumption~B, that the reconciliation is a useful approximation because it works. (In a little more detail, the density matrix for a classical world can be placed in diagonal form: Schroedinger's cat is either alive or dead. The density matrix for our quantum word has off-diagonal elements representing entangled states, and predicts expectation values and correlations of observables. A macroscopic object is entangled an immense variety of ways with observables in the system and the external world that produce myriad phase shifts that ``surely'' average out to zero in the off-diagonal elements in the macroscopic limit. Thus a molecule passing through the double slit experiment is in a coherent state in two wave packets that pass through the slits and then merges into the same molecule that leaves a macroscopic signature at the detector. This is sensible, consistent with practice, but hard to prove.)  

We cannot prove that research in fundamental physics is leading to an ever better approximation to some definite law of physics to be discovered; we must state it as Assumptions~C and~D. In the philosophy literature this amounts to the Duhem-Quine thesis that it is impossible to prove that our canonical physics is the only way to fit the available empirical evidence and maybe as well evidence we have not thought to seek. This is clear enough; we cannot judge the empirical basis for a theory we don't know, and we cannot disprove its existence. The American philosopher of science Norwood Russell Hanson (1961, p. 36) offered another way to think of the Duhem-Quine thesis.  
 \begin {quotation}
\noindent Given the {\it same} world, it might have been construed differently. We might have spoken of it, thought of it, perceived it differently. Perhaps facts are somehow moulded by the logical forms of the fact-stating language. Perhaps these provide a `mould' in terms of which the world coagulates for us in definite ways.
\end{quotation}
We must live with this. No matter how tight and accurate the agreement of theory and practice it remains conceivable that a different theory would do as well or even better. We can only aim to add tests of predictions that if successful make this unlikely thought even more unlikely, and add to the weight of evidence that we are approaching what we term reality. 

Thoughts along the lines considered here are expressed in the books {\it Doubt and Certainty} by the physicists Tony Rothman and George Sudarshan (1998) and {\it Dreams of a final theory} by the physicist Steven Weinberg (1992).

\subsection{Predictions and falsifications}\label{Predictions and falsifications}

Assumption~B in Section~\ref{sec:startingassumptions} express a central concept of science: a theory is to be judged by the degree of consistency of its predictions with the empirical evidence. If a prediction, something that was not anticipated, is observed, it is taken as evidence that the theory might be a useful approximation to the nature of the world. Why else the consistency? Maybe only an accident? Checking that calls for more predictions to be tested. 

Karl Raimund Popper's contribution to the philosophy of science is informally honored by physicists by the use of Popper's word, ``falsifiable,'' as in a list of advantages of a proposed theory that ends with `` and it is falsifiable." Popper's (1959, p. 10) more nuanced argument includes this: 
\begin{quote}
\noindent singular statements --- which we may call `predictions' --- are deduced from the theory \ldots those are selected which are not derivable from the current theory, and more especially those which the current theory contradicts. Next we seek a decision as regards these (and other) derived statements by comparing them with the results of practical applications and experiments \ldots if the conclusions [predictions] have been falsified, then their falsification also falsifies the theory from which they were logically deduced.
\end{quote}
The concept of empirical tests of predictions is essential to Assumptions~B and~C. People can be wonderfully imaginative in constructing stories, or theories, that account for what is thought to be appropriate. It is encouraging if a theory can be devised to fit many known observations, but that might be only a clever contrivance. That is why we rely on tests of predictions: implications that were not considered in constructing the theory. The greater the number and variety of successful predictions the better the evidence for the theory, and the more persuasive the case that the theory is a useful approximation to this aspect of reality. (But bear in mind that physical theories tend to have parameters that can be adjusted to fit the data. An adjustment of a parameter within the bounds allowed by other tests that improves the fit is encouraging but not a demanding check. The tests of non-negotiable predictions that remain after adjustments of such free parameters are crucial.)

Another way to put the situation is the philosophers' ``miracle argument'' (Putnam 1975, p. 73), that  
\begin{quotation} 
\noindent The positive argument for realism is that it is the only philosophy that doesn't make the success of science a miracle.
\end{quotation}
I take Putnam's ``success'' to be the many confirmations of scientific predictions in controlled experiments and observations as well as in practical applications. These successful applications are predictions, not likely to have been imagined by the people who devised the theories. Consider the many applications  of electromagnetism and quantum physics.  Maybe some successful predictions are only accidental, and maybe others are successful because many theories were examined and one at last found that happens to fit the measurements. But such thoughts are difficult to maintain for the numerous successful predictions of canonical physics reviewed in Section~\ref{sec:tests}. 

The American philosopher Charles Sanders Peirce made perceptive comments about this and other aspects of physics, much of it published in the magazine {\it Popular Science Monthly}. (The magazine was renamed {\it Popular Science}. I and I imagine many other youths read it and some became physicists in part because of that magazine.) The Canadian philosopher Cheryl Misak (2016), in {\it Cambridge Pragmatism From Peirce and James to Ramsey and Wittgenstein}, gives an informative account of Peirce and like thinkers. 

Peirce (1878, p. 299) presented a good illustration of Assumptions~C and~D:
\begin{quotation} 
\noindent all the followers of science are fully persuaded
that the processes of investigation, if only pushed far enough, will
give one certain solution to every question to which they can be applied. 
One man may investigate the velocity of light by studying
the transits of Venus and the aberration of the stars; another by the
oppositions of Mars and the eclipses of Jupiter's satellites; a third by
the method of Fizeau \ldots\ [and another] may follow the different methods of comparing the measures
of statical and dynamical electricity. They may at first obtain different results, but, as each perfects his method and his processes, the results will move steadily together toward a destined centre.
\end{quotation}
The first two of the measurements Peirce mentioned are from astronomical observations that use trigonometry and Newtonian  mechanics to relate the measured radius of the Earth to distances to planets, moons, and meteorites. The next is from laboratory measurements of the speed of light, and the last in this abbreviated quotation is from the theories of electricity and magnetism, which scientists were starting to realize predict the speed of electromagnetic radiation, as in light.  These are quite different ways to measure the speed of light. We can put it that one of these measurements produces a prediction, and the others are quite different methods of testing the prediction. The consistency is notable. Why should the motion of light coming to us from the planet Jupiter be the same as the speed of light on Earth and the speed derived from experiments on the properties of electricity and magnetism? The physicists' answer is that this is a demonstration of the unity of physics, evidence that the speed of light is a universal physical constant.

Peirce was confident that all would grow ever closer to agreement as the measurements improved; it was Peirce's understanding of reality. More examples are reviewed in Section~\ref{sec:tests}. They agree with the intuitive feeling of natural scientists that their research is establishing aspects of reality.

The four assumptions in Section~\ref{sec:startingassumptions} do not include Popper's falsifiability. To see why consider the classical theory of electromagnetism that Maxwell put together in the mid-1800s. By the 1930s Maxwell's theory had passed many tests by its applications in the laboratory and to a broad range of technology, things like transoceanic telegraph cables and electric streetcars. The theory nevertheless was known to fail when applied to an atom. One could say that this falsified Maxwell's equations, but a better way to put it is that the classical theory is a demonstrably useful approximation that requires improvement. That proves to be quantum electrodynamics, QED. It contains the classical theory as a limiting case,\footnote{I take it here and in all this essay that the conventional quantum measurement theory mentioned in footnote 2 in Sec.~\ref{sec:ruleslogic} is an adequate approximation.} as a viable theory must, and it yields new predictions that pass experimental tests, some to considerable accuracy, as an established theory should. This is reviewed in Section~\ref{sec:tests}. 

 But QED also is an approximation, a limiting situation in the electroweak theory that is part of the standard model of particle physics that the community agrees calls for improvement by discovery of another step toward the unification of science. This is Assumptions~C and~D.

\subsection{Physical reality}\label{sec:reality}

Physicists replace the rich array of philosophical and cultural ideas about reality with the operational approach in Assumption~D in Section~\ref{sec:startingassumptions}. Our physical world is assumed to be mind-independent and capable of being probed to reveal the rules, the fundamental theories, by which the world operates. It must be in successive approximations because tests have limited accuracy and probe only parts of the world. Hermann Bondi (1977) put it that ``science is only possible because one can say something without knowing everything.'' And we hope new approximations can be found and empirically shown to be better. Other scientists must have stated similar thoughts, I suppose many times, but I have not found other references.

We do not know there is a physical reality to be found, of course. Peirce  (1877, pp. 11, 12) wrote about this in a {\it Popular Science Monthly} article:
\begin{quotation}
\noindent Such is the method of science. Its fundamental hypothesis, restated in more familiar language, is this: There are real things, whose characters are entirely independent of our opinions about them; those realities affect our senses according to regular laws, and, though our sensations are as different as are our relations to the objects, yet, by taking advantage of the laws of perception, we can ascertain by reasoning how things really are \ldots\ It may be asked how I know that there are any realities \ldots If investigation cannot be regarded as proving that there are real things, it at least does not lead to a contrary conclusion.
\end{quotation}
This is sensible by physicists' way of thinking. Assumption~D is the expectation that reality is to be discovered, but of course we can't know that for sure, we can only say that our experience so far is encouraging.  Percy Bridgman (1927), the 1949 Nobel laureate honored for research on the behavior of matter at high pressure, offered a similar thought.
\begin{quotation}
\noindent It is of course the merest truism that all our experimental knowledge and our understanding of nature is impossible and non-existent apart from our own mental processes, so that strictly speaking no aspect of psychology or epistemology is without pertinence \ldots\ We shall accept as significant our common sense judgment that there is a world external to us, and shall limit as far as possible our inquiry to the behaviour and interpretation of this ``external'' world.
\end{quotation}
Research in natural science has advanced a lot since these remarks but they remain good expressions of implicit and sometimes explicit thinking in the natural science community. We have no evidence against the four assumptions in Section~\ref{sec:startingassumptions}, and considerable evidence for them from the successes of science. 

An interesting example of an alternative way of thinking about reality came from Bruno Latour's experience during two years embedded in the Salk Institute for Biological Studies. Latour knew nothing about natural science; he was in effect an anthropologist observing a previously unknown culture. It left him with mixed feelings about the reality of biological molecules. In the book about this remarkable experience Latour and Woolgar (1986) reported that (p. 183) 
\begin{quotation}
\noindent We are not arguing that [the biophysical molecule] somatostatin does not exist, nor that it does not work, but that it cannot jump out of the very network of social practice which makes possible its existence.
\end{quotation}
So is the biophysical molecule somatostatin real? 

To keep it simple let us consider first atomic helium. We have an excellent case for its reality. Helium emission lines are observed in plasma around the sun and in the plasma around massive stars in galaxies near and far. Its effect as a low mass density noble gas is required to account for the mass density as a function of radius in the theory of the structure of the sun, with predictions that are well tested by heliosismology (Aerts,  Christensen-Dalsgaard, and Kurtz 2010 Chapter~1). The same low mass density is tangible in a helium-filled balloon you hold by a string. The phase diagrams of condensed helium at low temperature in its two isotopes illustrate remarkable aspects of quantum physics. The evidence reviewed in Section~\ref{sec:physics} is that the energy levels of an isolated helium atom follow from quantum physics (as illustrated in eq.~[\ref{eq:helium}]). This adds to the evidence that helium atoms exist and quantum physics provides an excellent approximation to their structure. Much the same evidence applies to the next most complicated atom, lithium (eq.~[\ref{eq:lithium}]). All of this argues that the standard physical pictures of atomic helium and lithium are good approximations to reality. But quantum physics is not a useful tool for the study of the therapeutic effect of lithium. From the point of view of physicists lithium in a living body is real, along with all the other atoms that are detected by nuclear magnetic resonance measurements, but analysis of the effect lithium has on how people feel is far too complex for computation from first principles. Physicists must leave this emergent phenomenon in more capable hands.

Latour observed studies of the biological molecule somatostatin. In its less massive form (fourteen amino acids) it consists  of 219 atoms with 870 electrons. A computation from first principles of the properties of this molecule in isolation (assuming it can exist in isolation as a close to pure state), as has been done for atomic helium and lithium, is unthinkable. It would have to be done in successive approximations by successively improved models. The evidence reviewed in Section~\ref{sec:physics} is that the properties of this molecule follow from quantum physics, but a thorough check for somatostatin is a serious challenge. I return to this point in Section~\ref{sec:final theory}.

The rules and logic mentioned in Assumption~A are supposed to apply whether or not we are present to attempt to discover them. Peirce put it that ``real things \ldots\ are entirely independent of our opinions about them.'' We have a good case for real things. The galaxies of stars certainly look real, and the evidence is that the older stars in our galaxy were present and evolving well before people or any other kind of life capable of it could have been  looking at them. It seems equally sensible to accept observations of the spectra of radiation from distant galaxies have the familiar absorption lines of the chemical elements, but redshifted, to mean the chemical elements existed before there was life on Earth. But of course we cannot hope to send a robot to a planet in a distant galaxy to conduct chemistry experiments and report back. A term for this situation is ``objective reality,'' but on the advice of David Hogg (NYU) I use ``mind-independent reality.''

Let us pause to consider that fascinating things surely are happening on the surfaces of the immense numbers of planets around stars in the immense numbers of observable galaxies, things that are real in the mind of a scientist, and surely objectively real, but we will never observe. These are examples of loose ends in science.

\subsection{The place of physics in the natural sciences}\label{sec:physics}

What is the place for fundamental physics among the many lines of enquiry in the natural sciences? All seek accurate and useful accounts of Nature by research shaped by the operating conditions, some qualitative, some informative descriptions, some tightly constraining mathematical theories. Fundamental physics is the reductionist search for a mind-independent reality defined and constrained by the four assumptions in Section~\ref{sec:startingassumptions}. It aims for theories that yield reliable quantitative predictions that can be compared to reproducible and accurate measurements. This seriously limits the reach of fundamental physics but it aids the approach to what we are assuming is mind-independent reality. 

What is the relation to chemistry? Baird, Scerri, and McIntyre (2006) put it that
\begin{quotation}
\noindent Chemistry does sit right
next to physics, with all its lovely unifying and foundational theory. Squinting our
eyes up tight, it is possible to see chemistry as complicated applied physics. Even in
denial, we say we are materialists, but the material world of our denial is the foundational
world of physical theory, and so chemistry---in principle anyway---must be
reducible to physics. But this has never been much more than an article of faith.
\end{quotation}

\noindent Physicists tend to take it without question that chemistry could be derived from standard quantum theory if only chemistry were not so complicated. We have some justification. Molecular hydrogen is simple enough for a clean close to first principles quantum computation of its binding energy: the work required to pull apart a hydrogen molecule into two hydrogen atoms at rest. We see an example of an entangled quantum state in the ground level of an ammonia molecule, where the nitrogen atom is on both sides of the triangle of the three hydrogen atoms. This is contrary to usual logic but a consequence of the quantum physics discussed in Section~\ref{sec:ruleslogic}. In the first excited level the nitrogen atom in effect oscillates between positions on either side of the triangle. The energy released by the transition from this inversion oscillation to the ground level was used to power the microwave radiation in early generations of atomic clocks. Steps to the application of physics to chemistry have gone still further with the application of advances in computation and data storage, but Earman and Roberts (1999) put the broader situation correctly:
\begin{quotation}
\noindent The concept of a law of nature seems to us to be an important one for
 understanding what physics is up to, but it is a misguided egalitarianism 
 that insists that what goes for physics goes for all the sciences. \label{page:botany}
\end{quotation}

Philip Warren Anderson (2011, pp. 144, 201) put it more explicitly: \label{page:Anderson}
\begin{quotation}
\noindent The important lessons to be drawn are two: 1) totally new physics can {\it emerge} when systems get large enough to break the symmetries of the underlying laws; 2) by construction, if you like, those {\it emergent properties} can be completely unexpected and intellectually independent of the underlying laws, and have no referent in them \ldots\ I think almost all the things worth studying are irreducibly complex [and that] requires research which I think is as fundamental in its nature as any other.
\end{quotation}
Anderson was a good physicist with a good point to make. As Earman and Roberts wrote, Anderson's subject, condensed matter, is as intellectually interesting and challenging as elementary particle physics. What is more, research in condensed matter physics costs far less and contributes far more to the world economy. And there are far more irreducibly complex things, from condensed matter to botany to people, than phenomena that are useful probes of foundational reality.

Anderson's point suggests the thought that, just as unexpected properties of condensed matter physics such as superconductivity are emergent from quantum physics, quantum physics might be an effective theory emergent from something even more fundamental. There is a difference: condensed matter physicists and botanists must take account of the atoms and electrons and electromagnetic field that are actors in the best approximation to fundamental theory so far, while there is little manifest evidence of actors from some deeper level. There are hints from theory. The gauge/gravity duality Juan Maldacena (1999) conjectured has led to thoughts such as that expressed by Nima Arkani-Hamed (2012) that
\begin{quotation}
\noindent Attempts to make sense of quantum mechanics and gravity at the smallest distance scales lead inexorably to the conclusion that space-time is an approximate notion that must emerge from more primitive building blocks.
\end{quotation}
Particle physicists are quite aware of the challenge of unifying our micro- and macro-physics, and the possibility that this might lead to a deeper level of fundamental physics to be explored and established. 

I became interested in the topic of this essay too late to discuss it with Phil Anderson, who was a colleague at Princeton and an excellent physicist. I expect Phil would have insisted that I have not given proper credit to the study of complex systems. We agree on the science but differ on priorities. My essay is on the philosophy of fundamental physics summarized in the four assumptions in Section~\ref{sec:startingassumptions}. The philosophy of the science of complex systems is in no way inferior to that of fundamental physics, but it belongs in a different essay.

\subsection{The anthropic principle}\label{sec:anthropic}

The starting assumptions in this essay do not mention mathematics, the language of physics. That is because there are other ways to probe the world; consider Faraday's great discoveries of the properties of electricity and magnetism in descriptive terms. The physics community is divided about another way to probe the universe, the anthropic principle.

Brandon Carter (1974) introduced his thinking about this with the remark that
\begin{quotation}
\noindent Copernicus taught us the very sound lesson that we must not assume gratuitously
that we occupy a privileged {\it central} position in the Universe. Unfortunately there has
been a strong (not always subconscious) tendency to extend this to a most questionable dogma to the effect that our situation cannot be privileged in any sense. This
dogma (which in its most extreme form led to the `perfect cosmological principle' on
which the steady state theory was based) is clearly untenable, as was pointed out by
Dicke \ldots 
\end{quotation}
Dicke (1961) drew attention to the consistency condition that the time elapsed since the early stages of expansion of the universe had to have been long enough to have allowed stars to produce the chemical elements we need, but not so long that stars suitable for supporting our existence would have exhausted their supplies of nuclear fuel and faded away. The evolution ages of the oldest stars and the radioactive decay age of the solar system are consistent with these conditions, as is the time elapsed to a  reasonable present mean mass density in an expanding relativistic universe. It would have been a serious problem otherwise. 

Weinberg (1989) discussed another consistency condition. The cosmic mean mass density expected from quantum physics, if represented by Einstein's cosmological constant, $\Lambda$, is far larger than what is allowed by the standard relativistic cosmology. Nima Arkani-Hamed (2012) put it that
\begin{quotation}
\noindent This is the largest disagreement between a ``back of the envelope'' estimate and reality in the history of physics---all the more disturbing in a subject accustomed to twelve-decimal place agreements between theory and experiment.
\end{quotation}
\noindent What is more, we have good evidence that we flourish just as the rate of expansion of the universe is making the transition from slowing due to the attraction of gravity to increasing due to the effect of a positive value of $\Lambda$. Why should that be? Weinberg offered an anthropic explanation. If $\Lambda$ had been positive and much larger than observed the universe would have been expanding too rapidly to have allowed the formation of the galaxy of stars within which the chemical elements we need were produced by recycling  of matter through stars. If $\Lambda$ had been negative and large the universe would have collapsed before natural evolution could have produced observers such as us. So Weinberg invited us to imagine a statistical ensemble of universes, a multiverse, with $\Lambda$  different in different universes and typically large, whether positive or negative, as might be expected from the large value suggested by quantum physics. Then we would have expected to have flourished in a universe in the multiverse with the largest absolute value of $\Lambda$ allowed by our existence. This is at least roughly what is observed. 

A multiverse of universes is expected in some versions of the cosmological inflation picture of what our universe was doing before the very early stages of expansion. A test of sorts would be to compare the value of $\Lambda$ derived from cosmology to the range of values expected in those universes in a multiverse that have physics capable of supporting life of a kind that would take an interest in the value of $\Lambda$. But we do not have an adequate theory of the properties of universes in a multiverse, or of the kinds of physical theories that allow the formation of entities that take an interest in the value of $\Lambda$.

If research continues to fail to reconcile the value of $\Lambda$ from quantum physics with the value from cosmology we can anticipate formation of two camps.  Physicists on one side will insist on working even harder to avoid the anthropic argument, risking a lot of time spent looking for an improved theory with no guarantee of success. The other will accept the idea of a multiverse and the mixed feelings about it. On the positive side Polchinski (2019), a respected authority, wrote that
\begin{quotation}\noindent
there is no rational argument that a multiverse does not exist, or even that it is unlikely. 
\end{quotation}
The multiverse offers a way to account for the value of  $\Lambda$ from cosmology, but at the cost of a troubling departure from the test of physics by its predictive power. It is a resolution of a puzzling situation in physics with no empirical test. My faith in human ingenuity is such that I would not be surprised if a more elegant theory were found to resolve the problem with $\Lambda$. Maybe it's too much to hope that the theory would be falsifiable.

My former colleague John Archibald Wheeler liked new ideas. He encouraged  Carter's thinking about the anthropic principle and offered his own adventurous ``participatory anthropic principle,'' that we are entangled with the rest of the universe. This is standard quantum physics, but I cannot see how it could be relevant to Schr\"odinger's cat in the quick and dead states because the entanglement of the two states of the cat with all that is around us is so broad that I expect entanglement averages out to zero, leaving a diagonal density matrix, the cat in one of two classical states. I expect entanglement across the universe is even more suppressed. But progress demands adventurous ideas that might aid discovery of the fundamental reality postulated in the starting assumptions presented in Section~\ref{sec:startingassumptions}, or maybe demonstrate that the assumptions are inadequate in some sense. It is a great adventure to explore adventurous ideas or seek other new ways of thinking, but usually best to keep your day job.

\subsection{Philosophies of physics}
\label{sec:philosophies}

Physicists do not tend to see much use for philosophy. Weinberg (1992) expresses the feeling in the chapter {\it Against Philosophy} in his book, {\it Dreams of a Final Theory}:
\begin{quotation}
\noindent Physicists do of course carry around with them a  working philosophy. For most of us, it is a rough-and-ready realism [but]  we should not expect [philosophy] to provide today's scientists with any useful guidance about how to go about their work or what they are likely to find.
\end{quotation}
This makes sense; just let physicists get to work. But learned people, nonscientists and scientists, can offer thoughts that help us better understand what we are doing, maybe particularly so when disagreements force us to think about why we disagree. 

I admire Ernst Mach's (1902) book, {\it The Science of Mechanics}, for its informative discussions and elegant demonstrations of classical mechanics. Mach's demonstrations still serve as valuable teaching tools; I used them in my introductory physics lecture demonstrations. Mach was a good physicist and we should pay attention to his opinions.

To Mach the science of mechanics is an economical way to state empirical results. Mach wrote that the \label{auxiliaryconcepts}
\begin{quotation}
\noindent atomic theory plays a part in physics similar to that of certain auxiliary concepts in mathematics; it is a mathematical {\it model} for facilitating the mental reproduction of facts. Although we represent vibrations by the harmonic formula, the phenomena of cooling by exponentials, falls by squares of times, etc., no one will fancy that vibrations {\it in themselves} have anything to do with the circular functions, or the motion of falling bodies with squares. It has simply been observed that the relations between the quantities investigated were similar to certain relations obtaining between familiar mathematical functions, and these {\it more familiar} ideas are employed as an easy means of supplementing experience.
\end{quotation}
Mach's negative thinking about atoms could not last. Rutherford, Boltzmann, Einstein and others were using the model of atoms to arrive at predictions that were encouragingly similar to observations

Mach's auxiliary concepts still are essential parts of physics. Quantum operators on state vectors in an abstract space are used to compute wonderfully precise and successful predictions. The dark matter of physical cosmology is not directly detected at the time of writing, and for all we know will prove to be observable only by the effect of its gravity. If so dark matter might remain classified as another of Mach's auxiliary concepts, but the web of indirect evidence from the effects of its gravity is tight enough that dark matter has a place in standard and accepted physics.

Mach understood what we would term the predictive unifying power of Newtonian mechanics, as we see in these remarks:
\begin{quotation}
\noindent The riddle of the tides, the connection of which with the moon had long before been guessed, was suddenly explained [by Newton's theory] as due to the acceleration of the mobile masses of terrestrial water by the moon \ldots The trade-winds, the deviation of the oceanic currents and of rivers, Foucault's pendulum experiment, and the like, may also be treated as examples of the laws of areas [conservation of angular momentum].
\end{quotation}
In present-day thinking it is impressive that the compact formulation of Newtonian physics accommodates this broad list of phenomena and more: Mach could have mentioned the motions of the planets, their moons, and falling apples. The broad success of Newtonian physics offers an excellent case for the assumptions in Section~\ref{sec:startingassumptions}: Nature does seem to operate by rules we can discover. But Mach saw the situation differently. He wrote that
\begin{quotation}
\noindent It is the object of science to replace, or {\it save}, experiences, by the reproduction and anticipation of facts in thought. Memory is handier than experience, and often answers the same purpose. This economical office of science, which fills its whole life, is apparent at first glance; and with its full recognition all mysticism in science disappears.
\end{quotation}
Mach's ``anticipation of facts'' can be read to mean ``successful predictions,'' but he does not seem to have been interested in this way of putting it. Mach's thinking about the nature of the physics he understood so well is a mystery that has given his  philosophy of positivism a bad name.

The physicists' philosophy summarized in the four assumptions in Section~\ref{sec:startingassumptions} is an approximation to the philosophers' realism. Anjan Chakravartty (2017) wrote that 
\begin{quotation}
\noindent Scientific realism is a positive epistemic attitude toward the content of our best theories and models, recommending belief in both observable and unobservable aspects of the world described by the sciences. \ldots\ It is perhaps only a slight exaggeration to say that scientific realism is characterized differently by every author who discusses it.
\end{quotation}
This is a reasonable description of physicists' ways of thinking, though I would rather not mention ``belief.'' Physicists can adjust to physical evidence. Cheryl Misak (2016) considers Charles Sanders Peirce to be a  pragmatist who starts with this realism. Since Peirce demonstrated admirable understanding of physics (discussed in Secs.~\ref{Predictions and falsifications} and~\ref{sec:reality}) I take it that physicists have a pragmatist philosophy. It fits the fact that empirical tests of physical theories cannot check all eventualities to all accuracy, meaning physical theories cannot be empirically established as mathematical theorems. We must instead rely on pragmatic judgements of how well predictions fit evidence and what that signifies. David Hogg (2009) puts it that the acceptance of an advance in physics is a plausibility argument. This is accurate, but as Hogg points out many aspects of physics have become plausible enough to inspire confidence, though never absolute belief, in the thought that we are approaching reality. 

The discovery of the uncertainty principle in quantum physics was a surprise that still exercises philosophers and physicists (e.g. Mermin 2019), for good reason (as discussed in Sec.~\ref{sec:ruleslogic}). How could a particle not have a definition position? But although quantum measurement theory moves the search for the nature of reality farther from our common experience, and makes it more interesting, it fits the fundamental starting assumptions in Section~\ref{sec:startingassumptions}. That is, the physicists' working philosophy as I state it in Assumptions~A to~D is pragmatic, not  disturbed by quantum physics. 

Niels Bohr (1925) presented a review of the many phenomena that are suggestive of quantum aspects of physics. It is an excellent illustration of the rich empirically-driven side of the invention of quantum theory. In this paper Bohr's early statement of his correspondence principle is that the 
\begin{quotation}
\noindent demonstration of the asymptotic agreement
between spectrum and motion gave rise to the formulation
of the ``correspondence principle,'' \ldots [which] expresses the tendency to utilise in the systematic development of the quantum theory every
feature of the classical theories in a rational transcription
appropriate to the fundamental contrast between
the postulates and the classical theories.
\end{quotation}
This is an outline of a prescription for the algebra of quantum observables and the always serious condition that the predictions of the quantum theory agree with the classical theory in conditions where the classical version is known to be accurate. (Bohr's later thoughts about the complementary roles of observables that do not commute need not be considered here.) 

\subsection{Thomas Kuhn}\label{sec:Kuhn}

Why do scientists accept Assumption~A in Section~\ref{sec:startingassumptions}, that what is observed by eye and instruments that probe the world on scales large and small operates by rules we can discover? Thomas Kuhn's thinking about this is interesting because he was trained as a physicist at Harvard University but became skeptical of the physicists' philosophy and expressed his thoughts in the influential  book,  {\it The Structure of Scientific Revolutions} (Kuhn 1962, 1970a). Evidence of its influence includes Kaiser's (2016) report that the book had ``Cumulative sales [that] exceed one million copies, and at least sixteen foreign-language translations have been published,'' and the sociologist Andrew Abbott's (2016, p. 168) report of some 17,000 citations in the Web of Science in the humanities and social sciences in publications up to 2012, a mean rate of about one citation a day. Paul Hoyningen-Huene (1993) presents a close analysis of {\it The Structure of Scientific Revolutions} in the book {\it Reconstructing Scientific Revolutions}. Kuhn liked the analysis; he wrote in the forward that ``I recommend it warmly,''   
 
Kuhn's book, hereinafter SSR (the second edition), is particularly interesting to physicists because of Kuhn first-hand experience with physics. He studied physics when he was an undergraduate at Harvard University, graduated in 1943, then spent several years on war research tracking German use of metallic chaff to avoid radar detections, and then returned to Harvard and completed a doctoral dissertation (Kuhn 1949) on {\it The Cohesive Energy of Monovalent Metals as a Function of Their Atomic Quantum Defects}. His advisor was Nobel laureate John H. Van Vleck. The thesis was followed by a paper with Van Vleck as co author, and then Kuhn's (1950) single-author paper on {\it An Application of the W.K.B. Method to the Cohesive Energy of Monovalent Metals}, published in {\it The Physical Review}, a well-respected journal. Here is a trained physicist whose thinking is widely influential and a good foil for  conventional thinking in the physics community. 

Thomas Kuhn (1977) recalled how his thinking was shaped by his attempt to understand Aristotle's ideas about what we now term mechanics, writing that
\begin{quotation}
\noindent Aristotle had been an acute and naturalistic observer \dots\ How could his characteristic talents have failed him so when applied to motion? How could he have said
about it so many apparently absurd things? And, above all, why
had his views been taken so seriously for so long a time by so
many of his successors? The more I read, the more puzzled I became.
\end{quotation}
In an interview by Sk\'uli Sigurdsson (1990) Kuhn recalled that
\begin{quotation}
\noindent What Aristotle could be saying baffled me at first, until---and I remember the point vividly---I suddenly broke in and found a way to understand it, a way which made Aristotle's philosophy make sense. [It] first got me onto the idea of gestalt switches and changes in conceptual frameworks, which was to show up in the {\it Structure of Scientific Revolutions} in 1962.
\end{quotation}
Kuhn (1977) explained that
\begin{quotation}\noindent 
Position itself was a quality in Aristotle's physics, and a body that changed its position therefore
remained the same body only in the problematic sense that the
child is the individual it becomes. In a universe where qualities
were primary, motion was necessarily a change-of-state rather than
a state.
\end{quotation}
Thus Kuhn concuded (in SSR p. 120) that
\begin{quotation}
\noindent  
Aristotle and Galileo both saw pendulums, but they differed in their interpretations of what they both had seen.
\end{quotation}

Kuhn's interest in Aristotle's way of thinking led him to a concept of physics that differs from the physics community in ways that are worth considering. Kuhn asked (in SSR, pp. 166, 171) two good questions:
\begin{quotation}
\noindent 
	1. Why should progress also be the apparently 
	universal concomitant of 
        scientific revolutions?

\noindent	2. Does it really help to imagine that there is some one full, 
	objective, true account of nature and that the proper measure  
	of scientific achievement is the extent to which it brings  
	us closer to that ultimate goal?
\end{quotation}
They are direct challenges to the assumptions in Section~\ref{sec:startingassumptions}. Indeed, natural scientists have not been issued a guarantee that the world acts on the basis of mind-independent rules that can be discovered in approximations that are converging toward ever better theory. Natural scientists rarely question these assumptions, in part because our instructors and their instructors seldom questioned them, but certainly also because we have been encouraged by generations of advances in physics that continue to yield new predictions that agree with improving observations. This is what would be expected if we were approaching an ultimate goal, our Assumption~D. 

How did Kuhn's experience in physics affect his thinking about these two questions? Galison (2016) and Kaiser (2016) report that inspections of Kuhn's notebooks do not offer much evidence of what Kuhn thought of his experience as a physics undergraduate and graduate student. So it is important that we have Kuhn's recollections ranging from his childhood to experiences at Harvard to life after SSR in ``an edited transcript of a tape-recorded three-day discussion---essentially an extended interview'' in Athens on October 19-21, 1995, by Aristides Baltas, Kostas Gavroglu, and Vassiliki Kindi, in the book Kuhn (2000). These authors combine expertise in physics, history, and philosophy. 

In the interview Kuhn (pp. 272, 273) recalled that, after completion of a reduced three-year undergraduate degree in physics at Harvard, employment in war research, and returning to Harvard, 
\begin{quotation}
\noindent I was finding it [physics] fairly dull, the work was not interesting \ldots\ I couldn't go back and sit still for that undergraduate chicken-shit and go on from there.  So, I decided I'm going to take my degree in physics. But it also was clear, and becoming increasingly clear, that I was not being very much fulfilled by my graduate physics teaching.
\end{quotation}
This feeling is to be contrasted with Kuhn's (p. 276) more positive recollection of what was leading up to SSR:
\begin{quotation}
\noindent I used to think \ldots I could read texts, get into the heads of the people who wrote them, better than anybody else in the world. I loved doing that. 
\end{quotation}
Kuhn expressed similar thoughts in an interview by Sk\'uli Sigurdsson (1990):
\begin{quotation}
\noindent I wondered whether a physics career was what I really wanted. I was
very conscious of the narrowing, the specialization required, and \ldots\ I was beginning to look for alternatives. No one of those seemed more attractive than the rest, until all of a sudden I was asked to assist President [James B.] Conant in teaching an experimental General Education course on the history of science, through readings of case histories. It
sounded like a pretty good idea; it would be a good experience, a chance to work
with the President of Harvard, and also my first exposure to history of science.
So I grabbed the opportunity and found it fascinating.
\end{quotation}

Kuhn's story seems familiar. I have observed that some of the best theoretically-inclined undergraduate students dislike the approximation methods used in fundamental physics: why don't we just go straight to the applications of the fundamentals? But we can't do that; we must rely on approximations. Maybe Kuhn distrusted the clever methods of approximation of quantum many-body physics required by his PhD dissertation on the physics of condensed matter. The condensed matter physicist Phil Anderson (2011, p. 38) put it that
\begin{quotation}
\noindent This process of ``model building,'' essentially that of discarding all but the essentials and focusing on a model simple enough to do the job but not too hard to see all the way through, is possibly the least understood---and often the most dangerous---of all the functions of a theoretical physicist.
\end{quotation}
Anderson was discussing the models used in the first successful theory of superconductivity. Kuhn's doctoral dissertation employed quantum model-building that was less abstract but still  employed clever, one might say adventurous, approximations to quantum physics. Kuhn might have had his experience in condensed matter in mind in the statement in SSR (p. 179) that
\begin{quotation}
\noindent normal puzzle-solving research [is] possible \ldots\
as consequences of the acquisition of the sort of paradigm that
identifies challenging puzzles, supplies clues to their solution, and
guarantees that the truly clever practitioner will succeed.
\end{quotation}
This is a reasonable description of research in condensed matter physics if the paradigm Kuhn mentioned is the invention by ``clever practitioners'' of ways to approximate what is expected from quantum physics and arrive at reliable and useful approximations that account for observations. But the approximations in condensed matter physics are bold and Kuhn could have wondered whether they were only contrived to get the wanted answers. Kuhn seems not to have accepted that the successful predictions were useful approximations to quantum physics, or to have appreciated the power of the physicists'  treasured concept of unification. The puzzle of superconductivity could not have been resolved  within classical Newtonian mechanics and Maxwell's electromagnetism. Quantum physics solved the puzzle, left the classical theories as useful limiting cases, and explained why the light emitted from a heated gas of the atoms of a particular chemical element is concentrated at discrete wavelengths that are quite specific to the element. This unification, which is a central goal of science, is discussed further in Sections~\ref{sec:tests} and~\ref{sec:unity}.

Kuhn was naturally disposed to question the physicists' notion of reality.  Kuhn (SSR p. 206) wrote that
\begin{quotation}
\noindent There is, I think, no theory-independent way to reconstruct phrases like `really there'; the notion of a match between the ontology of a theory and its ``real'' counterpart in nature now seems to me illusive in principle.
\end{quotation}
Wikipedia (\url{https://en.wikipedia.org/wiki/Ontology}, accessed 28 December 2023) informs us that, in
\begin{quotation}
\noindent metaphysics, ontology is the philosophical study of being. It investigates what types of entities exist, how they are grouped into categories, and how they are related to one another on the most fundamental level (and whether there even is a fundamental level).
\end{quotation}
Fundamental quantum physics does not seem to have a secure ontology. The quantum field operators in an abstract multidimensional space are not likely candidates for it; they are better described as examples of Mach's auxiliary concepts. They enable wonderfully precise agreements of predictions and measurements, but are they ``really there?'' It this question meaningful? Physicists tend to ignore both questions.

Kuhn was interested in the thinking of people doing physics, while I am interested in Kuhn's thinking from  a physicists' point of view. I am led to offer a personal thought. Among my early memories is of looking into an older sister's schoolbook and seeing an illustration of a compound pulley (technically a block and tackle). I thought that was neat and still do. Kuhn was born to write SSR. I was born to write a close examination of SSR from a physicist's point of view. It is not surprising that we differ on many issues and that I align my positions to the physics I know and love. 

\section{Empirical Tests of Physical Theories}\label{sec:tests}

The subject for this section is evidence for the four assumptions in Section~\ref{sec:startingassumptions}, that scientific research can yield persuasive evidence of useful approximations to a unique unified fundamental physics. The central idea, which is considered in Section~\ref{Predictions and falsifications}, is that if a theory produces successful predictions within the uncertainties of theory and practice it is evidence that the theory is a useful approximation to what we take to be fundamental reality. Physicists tend to be proud of how well this has succeeded, maybe even arrogant: Don't bother thinking about making a machine that violates local energy conservation. But there is a reason, the remarkably tight consistency of theory and practice when both can be reliably established, which covers a considerable variety of experience.  Here are examples.

\subsection{Tests of quantum electrodynamics}\label{sec:precisiontests}

When Thomas Kuhn was writing SSR, the foil for this essay, the tests of the quantum theory of electrodynamics, QED, were celebrated for their precision. The following illustrations of the state of this physics around this time is taken from the Brodsky and Drell (1970) review, {\it The Present Status of Quantum Electrodynamics}.

The magnetic dipole moments of the electron, muon, and tau particles are written as $g = 2(1+a)$, where $g=2$ follows from Dirac's equation and $a$ is the effect of the quantum interaction of the particle with the electromagnetic field and the other field operators of standard particle theory. Brodsky and Drell concluded that the best estimates of the theory and observation of the quantum contribution to the magnetic dipole moment of the  electron, $e^-$, in 1970 was
\beqa
&&\hbox{predicted: } a^- = 0.001\, 159\, 663,  \nonumber \\
&&\hbox{measured: } a^-  = 0.001\, 159\, 646.
\label{eq:egfactor}
\eeqa
For the positron Broadsky and Drell found
\beqa
&&\hbox{measured: } a^+  = 0.001\, 160.
\label{eq:e+gfactor}
\eeqa
The theoretical values of the electron and positron moments are the same. For the positive and negative muons, the more massive relatives of the electron, Broadsky and Drell gave
\beqa
&&\hbox{predicted: } a_\mu  = 0.001\, 165\, 87, \nonumber \\
&&\hbox{measured: } a_\mu = 0.001\, 166.
\label{eq:muon}
\eeqa
I have slightly spoiled the precision of the theory by simplifying to the fixed value of the fine-structure constant presented in Broadsky and Drell. 

 I have not found evidence that Kuhn was aware of this line of research, and if so how it affected his thinking discussed in Section~\ref{sec:Kuhn}. It certainly led physicists then as  now to  conclude that equations~(\ref{eq:egfactor}) to~(\ref{eq:muon}) with related measurements make an excellent case that this branch of physics, QED, is a good approximation to reality. To repeat, why else would theory and practice be so close? The position taken in a branch of sociology is discussed in Section~\ref{sec:SSK}.

Recent measurements of the electron magnetic moment and related physical constants are so precise that a statement of the comparison of theory and observation requires more discussion than is useful here. (Bear in mind that it is meaningless to think of the numerical value of the speed of light $c$ as a universal constant because the value depends on the choice of units. The use of a centimeter or an inch does not matter for most of the results discussed in this section, as long as the unit is stated. But the standard inch and centimeter are not defined well enough for the precision measurements of the magnetic dipole moment of the electron. Instead measurements are reduced to dimensionless quantities such as the fine structure constant $e^2/\hbar c$ that do not depend on standards of length, time, or mass. Arbitrary values of remaining free parameters enable statements in familiar units.) I instead offer the concluding statement by Fan, Myers, Sukra, and Gabrielse (2023):
\begin{quotation}
 \noindent The most precise prediction of the SM [standard model of particle physics] agrees
with the most precise determination of a property of an
elementary particle [the quantum contribution to the magnetic dipole moment of the electron with related measurements] to about 1 part in $10^{12}$.
\end{quotation}
The value of $a$ for a muon is a little different from the electron because the greater muon mass increases the effect of the quantum interactions with the other quantum fields. The present prediction from Aoyama et al. (2020) and measurement from Aguillard et al. (2024) are
\beqa
&&\hbox{predicted: } a_\mu  = 116\, 591\, 810.(43)\times 10^{-11}, \nonumber \\
&&\hbox{measured: } a_\mu = 116\, 592\, 059.(22)\times 10^{-11}. \label{eq:recent_muon}
\label{eq:muon}
\eeqa
The numbers in parentheses are the uncertainties in the last two digits. 

Theory and measurement for the muon differ by three or four  standard deviations. This might be a significant difference, and if so it might be an indication that the muon interacts with a field not yet detected and entered in the standard model for particle physics. Or maybe it indicates a flaw of some sort in standard quantum physics, but if so the flaw is tiny and not likely to be seriously considered until sources of error within the parameters of standard physics, including the possibility of a new field, have been thoroughly explored.

The measurement of the magnetic moment of the tau lepton, the still more massive relative of the electron, is really difficult, and not yet a demanding check of quantum physics and the standard model for particle physics.

Fan et al. (2023) state a motivation for these spectacularly accurate measurements: the ``quest to find physics beyond the Standard Model of Particle Physics.'' Physics is not complete and clues to how to improve it might be found in discrepancies between theories and measurements of $a^\pm$ for the electron, muon, and tau leptons. This is a common way of  thinking in all branches of science. A second motivation is the natural human impulse to aim for the best possible measurement. A third, or maybe an unintended consequence, is the demonstration of how close theory and practice can be in physical science. 

There were and are other precision tests of QED, but this illustrates the situation. Quantum Chromodynamics, QCD, the theory of the strong interaction, grew as a parallel to QED; it now passes the rich variety of tests reviewed by Campbell, Huston, and Krauss (2018) in {\it The Black Book of Quantum Chromodynamics: A Primer for the LHC Era}. The remarkable precision and consistency of theory and observation deeply impressed physicists in the 1960s and still does. 

\subsection{The quantum physics of atoms}\label{sec:atoms}

Another illustration of the powerful empirical support for Assumptions~B and~C in Sec.~\ref{sec:startingassumptions}) comes from the application of quantum theory to the structures of atoms. Computation of the structure of a helium atom its ground level requires serious numerical computation to take account of the two electrons bound to the more massive atomic nucleus, but important results were obtained using 1950s-era computers. Kuhn could have been aware of the celebrated paper, Pekeris (1959), which reported a precision computation of the ionization energy of atomic helium: the energy required to pull an electron from a helium atom in its ground level. The result compared to the measurement was
\beqa
&&\hbox{predicted:  } 198\, 310.687\hbox{cm}^{-1}, \nonumber \\
&&\hbox{measured: } 198\, 310.82\pm 0.15\hbox{cm}^{-1}.
\label{eq:helium}
\eeqa
The three electrons in the next heavier element, lithium, make a precision computation of its structure much more difficult, but it has been done more recently. King (1999) reported the lithium ionization potential 
\beqa
&&\hbox{predicted: } 43\, 487.14(2) \hbox{cm}^{-1},  \nonumber \\
&&\hbox{measured: } 43\, 487.150(5) \hbox{cm}^{-1}.
\label{eq:lithium}
\eeqa
It is natural to push on to similar computations of properties of berylium, boron, and on, with increasing difficulty. I discuss the importance of this research in Section~\ref{sec:final theory}.

These precision tests of energy levels finesse the challenge of unifying classical and quantum physics (the off-diagonal terms in the density matrix mentioned in footnote~2 in Sec.~\ref{sec:ruleslogic}) by focusing on good approximations to energy eigenstates, where the meaning of energy agrees well enough with classical physics. This standard physics yields the remarkably precise tests of consistency of theory and measurement in equations~(\ref{eq:helium}) and~(\ref{eq:lithium}). It is what would be expected if quantum physics is a useful approximation to reality, and it is an important illustration of what physicists have in mind in Assumptions~B and C.

\subsection{Measurements of the fundamental physical constants}\label{sec:FundamentalPhysicalConstants}

The check of consistency of results from different kinds of measurements of a physical constant, such as the speed $c$ of light, is a test of the theories used to interpret the measurements and a test of the idea that $c$ and the other physical constants have universal meaning. Peirce's (1878) explanation of this point is discussed in Section~\ref{Predictions and falsifications}.  Since then a great deal of effort has been devoted to obtaining the best possible measurements of $c$ and the other fundamental constants. They are important as standards,  it is a natural impulse to make the measurements as good as possible, and the results are important evidence of the reality of the speed of light and the unity of physics.

The physicist Raymond Thayer Birge (1929) reported three different measurements of $c$ (in units of $10^{10}\hbox{cm~s}^{-1}$):
\beqa
&& c = 2.99796  \hbox{ from laboratory timing of reflected light pulses},\nonumber\\
&& c = 2.9970 \hbox{ from wavelengths and frequencies of  standing  waves},\nonumber\\
&& c = 2.9971 \hbox{ from the ratio of electrostatic to magnetostatic units}. \label{eq:Birge29}
\eeqa
The treatment of measurement uncertainties was not well organized in 1929; uncertainties in these three quantities likely are in the last digit quoted. Peirce (1878) had mentioned results from the first and last of these methods along with several astronomical measurements that  could not be made accurate enough for Birge's purpose but remain important historical evidence that $c$ in the solar system agrees with terrestrial measurements, as would be expected of a measure of reality. 

Birge wrote that
\begin{quotation}
\noindent The decision as to the most probable value, at a particular time, of any
given constant, necessarily demands a certain amount of judgment \ldots
\end{quotation}
We have no fixed and final evidence in natural science. Of course Birge had to make judgement calls: which of the measurements in equation~(\ref{eq:Birge29}) is most trustworthy? Reputations of the authors are influential, though who knows how fair or unfair that is in each particular case? But let us bear in mind that the close agreement of the three numbers in equation~(\ref{eq:Birge29}), which were obtained in quite different ways by different groups, is serious evidence that they were obtained using theories that are close to the physics of reality. 

Birge reanalyzed data from Millikan's (1917) report of measurements of the electric charges on water and oil drops, and concluded that the electron charge is 
\beq
e= -(4.768\pm 0.005)\times10^{ -10})\hbox{ absolute electrostatics units},
\label{eq:Birge_e}
\eeq
with the statement that ``This, the writer believes, is the most reliable value that can be deduced from Millikan's oil-drop work.'' Millikan (1917) had reported $e=-4.774\pm 0.005$, close enough to Birge's value.  

In the book, {\it Scientific Knowledge: A Sociological Analysis}, Barnes, Bloor, and Henry (1996, pp. 33 - 45) reviewed evidence found in Millikan's surviving notebooks of charge measurements that Millikan discarded as faulty, maybe because of problems with experimental setups, perhaps because some measurements simply seemed questionable. It led some to wonder whether Millikan rejected measurements that indicated charges less than the theoretical minimum, the charge $e$ of a single electron. That is not something you would expect of such a capable and experienced physicist, and debate on this issue became known in some circles as the ``battle of the electron.'' It is notoriously difficult to control unconscious bias against unexpected results such as this even with serious attempts to control it. Maybe Millikan, despite the great care of a good physicist, overlooked a fractionally charged particle, maybe a quark if theory somehow allowed it. But all this is unlikely because Birge had a check of consistency from a different method, electrolysis.

Electrolysis gives the mass transported by the transport of charge that could be accurately measured: current times time. That with molecular weights and lattice spacings in crystals from X-ray diffraction gives a measure of $e$. Birge quoted results from this approach by two groups:
\beq
e=-4.776 \hbox{ and } e=-4.794, \label{eq:electrolysis}
\eeq
in the units in equation~(\ref{eq:Birge_e}). Birge placed greater trust in the first value. Both differ from Millikan's result by more than Millikan's stated uncertainty, it is thought due to an error in Millikan's viscosity of air. But they are close. Common experience is that actual errors tend to exceed estimates because subtle sources of uncertainty can be difficult to find.

Birge's values of $c$ (in {eq.~[\ref{eq:Birge29}]) and the values of $e$ (in eqs.~[\ref{eq:Birge_e}] and~[\ref{eq:electrolysis}]) are consistent within sensible allowance for systematic errors. Physicists take the fact of consistency of results derived from quite different ways to probe the world to be evidence that the world operates by rules that can be discovered. This is our Assumption~A. It is simple enough.

Richard Cohen and Jesse DuMond led a celebrated a series of papers in the mid-1940s to the mid-1960s on the values of the fundamental constants. DuMond and Cohen (1953) reported improved measurements
\begin{quotation}
\noindent chiefly made possible by the intense development
since World War II of microwave and atomic beam
techniques for the study of proton resonance in magnetic
fields, the fine structure of energy levels in hydrogen and
deuterium, the magnetic moments, spin gyromagnetic
ratios, nuclear magnetic resonance frequencies, and
cyclotron frequencies of fundamental particles such as
protons and electrons, etc.
\end{quotation}
Precision measurements by applications of this variety of technology to a variety of phenomena placed tight constraints on combinations of values of the fundamental parameters, including $c$ and $e$, always provided standard physics is a sufficiently accurate approximation to physical reality. DuMond and Cohen (1953) reported the values
\beqa
&& c=299792.9\pm 0.8\hbox{km sec}^{-1}, \nonumber\\
&& e= -(4.80288\pm 0.00021) \times 10^{-10}\hbox{ esu}. \label{eq:DuMondCohen}
\eeqa
The nominally most precise value for $c$ from Birge's 1929 list in equation~(\ref{eq:Birge29}) differs from DuMond and Cohen by one part in $10^5$. This is impressive;  $c$ certainly appears to be a physically real constant. Birge's estimate of $e$  (eq.~\ref{eq:Birge_e}) from Millikan's notebooks differs from the DuMond and Cohen value by $0.7\%$. The ``battle of the electron'' was worthwhile; near endless checks are worthwhile; but no flaw appears in Millikan's experimental result. The measurements of $e$ were not as precise as the measurements of $c$, but the charge measurements yield independent evidence of mind-independent reality.

To summarize, we have the sensible and reasonable degree of agreement of the speed of light that Peirce (1878) reported, the speed of light and the electron charge  Birge (1929) found, the DuMond and Cohen (1953) reports, and the recent CODATA (Committee on data of the international science council) recommended values of the fundamental constants of physics and chemistry presented in Table~31 in Tiesinga, Mohr, Newell, and Taylor (2021). There are apparent anomalies, notably the three or four standard deviations difference between the prediction and measurement of the magnetic dipole moment of the muon. It is judged to be likely significant and a hint to an addition to the particle physics model, always an interesting possibility. But the greater point is the consistency of results from the long tradition of precision measurements of $c$, $e$, and the other physical constants. The physics community treasures this evidence of the unity of physical science.
 
\subsection{Detection of the neutrino}\label{existence_tests}

Kuhn, our foil for physics' conventional ways of thinking, was aware of great developments in post-WW~II particle physics. He wrote (in SSR pp. 26, 27) of
\begin{quotation}
\noindent the gigantic scintillation counter designed to demonstrate the existence of the neutrino---these pieces of special apparatus and many others like them illustrate the immense effort and ingenuity that have been required to bring nature and theory into closer and closer agreement.$^3$ [Footnote 3 is a list of references to experiments, including that of Reines and Cowan to be discussed here.] That attempt to demonstrate agreement is a second type of normal experimental work, and it is even more obviously dependent than the first upon a paradigm. The existence of the paradigm sets the problem to be solved; often the paradigm theory is implicated directly in the design of apparatus able to solve the problem. Without the {\it Principia}, for example, measurements made with the Atwood machine would have meant nothing at all.
\end{quotation}
Indeed, without the theory detection of the neutrino would have been puzzling, and of course the experiment would be exceedingly unlikely to have been done. But physicists had a successful theory of nuclear reactions involving the creation and annihilation of electrons that required hypothetical particles, the neutrino and antineutrino. The direct detection of the antineutrino in the 1950s was an important addition to the case that the theory is a useful approximation to reality. Instead of a precise measurement this is the demonstration of existence of a hypothetical particle. 

Wolfgang Ernst Pauli introduced the idea of neutrinos to save local conservation of energy and momentum in nuclear reactions, and Enrico Fermi introduced the four-fermion theory of the interaction of a neutrino with an electron, a proton, and a neutron, as in the reactions 
\beqa
&& {\rm n} \rightarrow {\rm p} + \bar\nu + {\rm e}^-,\label{eq:npnue}\\
&& {\rm p} +\bar\nu \rightarrow {\rm n} + {\rm e}^+.\label{eq:nupne}
\eeqa
The proton and neutron are n and p, ${\rm e}^+$ is a positron, ${\rm e}^-$ an electron, and $\bar\nu$ an antineutrino. The first line describes the decay of a neutron into a proton with the creation of an electron that conserves electric charge and an antineutrino that conserves lepton number. In the second line the annihilation of an antineutrino by a proton with the creation of a positron and the conversion of the proton to a neutron follows by a rotation of the in and out states in the first line. It is the same quantum theory. 

Fermi's theory with refinements had served well in experiments in nuclear physics, but in 1950 the neutrino in these reactions was a postulate. Now dark matter in the standard cosmology is a postulate required by a successful theory but not detected at the time of writing. A serious difference is that the 1950 theory predicted that neutrinos could be detected, by a difficult but feasible experiment.

The first direct detections in the 1950s (Cowan, Reines, Harrison, Kruse, and McGuire 1956 and references therein) were of antineutrinos from a nuclear reactor, where fissions of neutron-rich heavy atomic nuclei released neutrons that decayed by equation~(\ref{eq:npnue}) and produced the antineutrinos that were streaming out of the reactor. A tiny fraction of them were annihilated in a detector near the nuclear reactor by the reaction in equation~(\ref{eq:nupne}), where the proton was in an atomic nucleus. The $\bar\nu$ annihilation would be accompanied by the creation of a positron that would be annihilated with an electron in the detector to produce gamma rays that could be detected. Some  neutrons could be captured by an atomic nucleus with the production of more gamma rays. The expected rate of $\bar\nu$ annihilations in the detector was computed from standard quantum physics using the laboratory measurement of the neutron half-life in the reaction~(\ref{eq:npnue}). 

Reines and Cowan designed their experiment to detect the flux of antineutrinos predicted by the theory. This is what Kuhn charged (in the quotation at the start of this section). But it is good physics, a test of a prediction. If antineutrinos had not been detected at the expected rate, within the uncertainties of the neutron half-life and what was happening in the reactor, the falsification would have been a serious challenge to standard physics. Why would the reactions in equations~(\ref{eq:npnue}) and~(\ref{eq:nupne}) have successfully fit nuclear reaction measurements involving electrons and positrons but failed direct neutrino detection? But the flux of antineutrinos was detected.

The physics community welcomed the Reines and Cowan detection, though not with the enthusiasm it merited because it was expected. The detection ranks in importance with the precision measurements reviewed in Sections~\ref{sec:precisiontests} to~\ref{sec:FundamentalPhysicalConstants} because it was a confirmation of a prediction that added to the weight to the evidence of the existence of neutrinos and the reliability of quantum theory. The Nobel Prize Committee recognized this some 40 years later.

\subsection{The general theory of relativity}\label{sec:GR}

Einstein's general theory of relativity that Kuhn encountered in the 1950s was a good example of a social construction: acceptance driven by elegance. When Kuhn was writing SSR the scant empirical support for Einstein's theory (detailed in Peebles 2022a, chapter 3) could lead a skeptic to say, ``but Einstein just made up this theory.'' I do not know how well Kuhn knew the state of tests of general relativity, but we have an indication that the empirical situation was even worse than Kuhn thought. 
 
 Kuhn wrote in SSR  (p. 155) that Einstein
\begin{quotation}
\noindent seems not to have anticipated that general relativity would account with precision for the well-known anomaly in the motion of Mercury's perihelion, and he experienced a corresponding triumph when it did so.
\end{quotation}
The triumph was limited. In a letter to Conrad Habicht in 1907 Einstein wrote that
\begin{quotation}
	\noindent At the moment I am working on a relativistic 
	analysis of the law of gravitation
	by means of which I hope to explain the still
	unexplained secular changes in the
	perihelion of Mercury [and added at the bottom of the page] so far, however, it does not seem to be going anywhere.
 \end{quotation}
(The English translation is in the Collected Papers of Albert Einstein, Vol. 5, Doc. 69.)
 
We see that the explanation of the anomaly was not a pure prediction: Einstein found what he was seeking. There is nothing wrong with this, of course, except that it reduces the significance of the agreement of the theory with the observations of the motion of the planet Mercury as a test of general relativity. 

When Kuhn was writing SSR there were three standard tests of general relativity. One, the explanation of the anomalous motion of Mercury, was important but we see that it is a dubious test. The second, the measurement of the gravitational redshift, was in a confused state. The white dwarf star Sirius~B was thought to offer the best test because its mass is comparable to that of the sun and its radius is much smaller, making the predicted gravitational redshift much larger than on the sun, and maybe more reliablly measured. St. John (1932) listed predicted and measured redshifts both equal to 19, in units that are not needed. This looked good. But Greenstein, Oke, and Shipman (1971) found the predicted value $83\pm 3$, because better estimates of the radius of Sirius~B are smaller, and they found measured value $89\pm 16$, partly because Sirius~B was farther from its much more luminous companion Sirius~A, making the light from this companion less distracting. The Greenstein et al. prediction and observation are consistent, and both are four times the consistent earlier estimates. There was no skulduggery; just honest attempts at difficult measurements and a willingness to accept what proved to be an erroneous estimate of the radius of Sirius~B, I suppose at least in part because the computed redshift seemed to agree with the measurement. That left one test of general relativity, the gravitational deflection of light by the mass of the sun. The deflection was detected, measured to a few tens of a percent by observations through the 1930s. It disagreed with the  estimate from Newtonian gravity theory, which was important, and agreed with the  prediction of the general theory of relativity (Trumpler 1956), which was important but slender empirical support for general relativity.

Janssen and Renn (2022) present a careful analysis of {\it How Einstein Found his Field Equations}. For our purpose it is sufficient to note that Einstein concluded that he had the right theory because of the compelling elegance of his 1915 field equations. To express the situation in modern terms, general relativity is the covariant classical  tensor field theory that preserves local conservations of energy and momentum with the simplest acceptable Lagrangian density, just as classical electromagnetism is the covariant classical vector field theory that preserves charge with the simplest acceptable Lagrangian density. This is why general relativity is presented with classical electromagnetism in the book, {\it The classical theory of fields}. My copy, Landau and Lifshitz (1951), is the English translation of the Russian 1948 edition. It is the second volume in a celebrated series, the {\it Course of Theoretical Physics}, which is a standard and respected compendium of theoretical physics. The presence of general relativity in this series is recognition that it was (and remains) a canonical part of theoretical physics, along with classical electromagnetism. But in the 1950s classical electromagnetism had been thoroughly tested by experiments and an enormous variety of  practical applications while general relativity had meagre support. Yet general relativity before 1960 was standard and accepted by the authority of respected theoretical physicists for its compelling mathematical elegance.

General relativity now passes demanding checks of predictions on a broad range of scales: tests of the inverse square law of gravity in the laboratory down to lengths of about 0.1~mm (Lee, Adelberger, Cook, et al. 2020); timing tests of electromagnetic signals on the scales of the Earth, $\sim 10^{7}$cm,  and the solar system, $\sim 10^{13}$cm, reviewed by Will (2014, 2018); tests of general relativity and the search for gravitational waves by precision measurements of periods of binary pulsars in the Milky Way galaxy, at distances $\sim 10^{22}$cm (Hulse and Taylor 1975; Will 2014, 2018; Agazie, Anumarlapudi, Archibald., et al., 2023); images of the shadow of the massive black hole in the center of the galaxy M~87 that is $\sim 10^{25}$cm from us (Event Horizon Telescope Collaboration 2019); detection of gravitational waves from merging black holes at distances $\sim 10^{27}$cm (LIGO Scientific Collaboration and Virgo Collaboration 2016); and the  cosmological tests reviewed in Peebles (2022a) that probe the universe at close to the largest detectible scale according to standard theory. 

The cosmological tests include precision measurements of the anisotropy spectra of the thermal cosmic microwave background radiation (CMB) temperature and polarization. The measurements show the patterns produced by oscillations of the interacting baryonic matter and radiation ending at redshift $z\sim 1000$, when the primeval plasma combined and released the radiation. This physical situation is simple enough to allow secure predictions of the effect of the matter-radiation coupling and decoupling, computed in reliable perturbation theory. And the microwave sky observed in space and favorable sites on Earth is clear enough to allow precise and accurate measurements. The theory and measurement of the power spectrum of the large-scale space distribution of the galaxies are less precise because there are far fewer galaxies than CMB photons, and the more complicated behavior of the baryons must be considered, but the oscillation signal is clearly detected and consistent with the signal in the CMB. This is important because the distributions of the CMB and the matter traced by galaxies evolved in different ways and are observed by quite different probes of the universe: the valued check of consistency from different methods and phenomena. The numerical values of the parameters in the standard cosmology are constrained by these measurements and other tests, typically to a few tens of percent. This is not very precise, but the evidence of consistencies of constraints on these parameters is important because it is derived from what was happening at quite different stages of expansion of the universe, ranging from the present back to the time of formation of the isotopes of hydrogen and helium when the mean distance between nucleons was nine orders of magnitude smaller than it is now and the temperature nine orders of magnitude larger.

There are hints of discrepancies of theoretical constraints on parameter values from the cosmological tests. That does not seem surprising because what behaves like dark matter and Einstein's cosmological constant in the present theory are modeled in simple ways that leave room for better theories. Also to be considered is that Einstein's field equation for general relativity and all the rest of standard physics might be a little different from what is found locally in the application to the immense scales of cosmology. There is ample time for this to happen. But easier to imagine is that the hypothetical dark matter and cosmological constant can be more realistically modeled. The present models seem far too simple, and there is empirical evidence to guide the search for something better.

The standard relativistic theory of the large-scale structure of the universe is incomplete. Notably, we do not have a tested theory of what the universe was doing before it was expanding. The tests we do have of macroscopic physics are far less thorough and precise than in the microscopic limit, and what is more general relativity with cosmology on large scales is not consistent with quantum physics with the standard model for particle physics on small scales. The conditions under which they have been tested so far are different enough that this has caused no problem with empirical tests, except for the puzzle of the quantum and cosmological vacuum energy densities. These are among the loose ends in fundamental physics. Some will be resolved, maybe not all. But the evidence I have mentioned is not likely to go away; we have a convincing case that relativistic cosmology isa good approximation to reality, which must be considered in the search for a deeper physical theory of reality.

\section{Sociology}  \label{sec:sociology}

I turn now to thoughts about physics to be found in sociology. Some of it makes good sense to me, some not.

Kuhn  wrote in SSR (p. 8) that ``many of my generalizations are about the sociology or social psychology of scientists.'' Consistent with this is Kaiser's (2016) report that, in the abundant correspondence Kuhn received about SSR, 28\%\ were about about psychology, sociology, or philosophy, and only 11\%\ were about  physics or chemistry. Physicists tended to be a little offended by Kuhn's suggestion that psychology and sociology play a role in the establishment of our supposedly objective physical principles, though we are people and it must be so. The question is the degree to which nominally well-established evidence and physical theories have been influenced by society, and how far that has confused and frustrated the attempts to find good approximations to physical reality, assuming there is reality. I offer examples of the interaction of science and society that seem relevant to physics.  

\subsection{Elegance}\label{sec:elegance}

Einstein offered the elegant thought that (in the English translation by Sonja Bargmann 1954)
\begin{quotation}
\noindent The supreme task of the physicist is to arrive at those universal elementary laws from which the cosmos can be built up by pure deduction.
\end{quotation}
\noindent
I cannot judge how serious Einstein was. Weinberg (1992), in his book {\it Dreams of a Final Theory}, was careful to explain that if we had the final physical theory, Einstein's  ``universal elementary laws,'' it would not improve weather forecasts; far too complicated. But a variant is attempted: deduce better approximations to the final theory by pure thought guided by known physics, with proper attention to elegance. This can produce good results.  Einstein accomplished it in his general theory of relativity. As we have seen in Section~\ref{sec:GR} this theory was for many years socially accepted, which does not establish it, but it now passes demanding tests that convincingly argue it is a useful approximation to reality. And now the physics of general relativity figures in the concept of a  next level of reality, a universe of universes, a multiverse. Some like this fascinating conjecture, some are not so sure. 

There are other arguments for and against elegance. The relativistic theory of the expanding universe (Sec.~\ref{sec:GR}) passes demanding tests, which is encouraging, but there is not a tested theory of the universe before it was expanding. The community favorite for the very early universe is cosmological inflation, which some argue is an elegant addition to cosmology inspired by quantum physics. It offers to  pushes back in time the question of what the universe was doing earlier than what follows from standard physics, but it leaves open the question: what was happening before inflation could apply? There is the concept of eternal inflation that would extend to the indefinite future, but it cannot have an eternal past because old universes would intrude. Einstein's cosmological constant is required in the standard cosmology but it is exceedingly inelegant because the only known way to reconcile its value with standard physics is the anthropic principle (Sec.~\ref{sec:anthropic}). Community feeling about that is mixed. The hypothetical cold dark matter of the standard cosmology is inelegant because it was introduced {\it ad hoc} to resolve a puzzle in cosmology, and it is detected so far only by its gravity. Maybe a consolation is that Maxwell had to introduce the hypothetical vacuum displacement current to find a unified theory of electricity and magnetism. Both hypotheses, displacement current and dark matter, now pass demanding tests of predictions

The Nobel laureate Paul A. M. Dirac is celebrated for his important contributions to quantum theory, and remembered too for his thoughts about the elegance and beauty of successful physical theories. An example is the statement that ``A physical law must possess mathematical beauty'' (often cited, and Dirac is said to have stated during a visit to the University of Moscow in 1956). Kuhn (in SSR p. 155) made the point more broadly, remarking on the influence of
\begin{quotation}
\noindent arguments, rarely made entirely explicit, that appeal to the individual's sense of the appropriate or the aesthetic---the new theory is said to be ``neater,'' ``more suitable,'' or ``simpler'' than the old.
\end{quotation}
Nobel laureate Frank Wilczek (2015) described his feeling of the elegance of fundamental physical theory in the book, {\it A Beautiful Question: Finding Nature's Deep Design}. Nobel laureate Steven Weinberg (1992) wrote in his book, {\it Dreams of a Final Theory}, that
\begin{quotation}
\noindent in this century, as we have
seen in the cases of general relativity and the electroweak
theory, the consensus in favor of physical theories has often
been reached on the basis of aesthetic judgments before the experimental
evidence for these theories became really compelling.
I see in this the remarkable power of the physicist's sense
of beauty acting in conjunction with and sometimes even in opposition
to the weight of experimental evidence.
\end{quotation}
Nobel laureate Roger Penrose, in an interview for a  documentary, declared that 
\begin{quotation}
\noindent I think beauty is a clear guide to truth \ldots\ And it's certainly the case that if you have two alternatives where you worry about which is true, it's a better bet to think that the one which is more beautiful is more likely to be true. But this is always a very subtle issue. You might find there's a deeper reason that you hadn't realised before which makes the other one actually beautiful in a deep sense that you hadn't appreciated before.
\end{quotation}
(This remark is from https://www.whyarewehere.tv/people/roger-penrose/ \copyright\ Ard Louis and David Malone.)

Just as elegance led the community to accept general relativity well before its great empirical successes some argue that superstring theory, which is an elegant extension of special relativity and the standard model for particle physics, might similarly be elevated to established theoretical physics on Richard Dawid's (2013) grounds of ``non-empirical confirmation,'' that is, by the consensus of experienced and respected physicists. This philosophy worked for general relativity, which was canonical physics before it passed demanding tests. Why not the same for superstring theory? 

Thoughts about beauty are shared by many, but Penrose's cautionary remark is appropriate. We have no  definition of beauty, and we cannot rely on beauty or elegance to assess physics because people are adaptable: we learn to like what is successful and might come to consider it beautiful. What is more, judgements from elegance can be wrong. The steady-state cosmology is elegant but it is convincingly ruled out. Most agree that Einstein's general theory of relativity would be more elegant without Einstein's cosmological constant, $\Lambda$, though the evidence requires it. (The convincing evidence for the presence of $\Lambda$ was established after Weinberg wrote his Dreams book.) These failures explain why Assumptions~A to~D in Section~\ref{sec:startingassumptions} do not mention beauty or elegance. 

But we cannot ignore the elegance of physical theory. I feel it, I expect many colleagues do too, and I am reminded of it when I hear exclamations such as ``isn't that neat.'' This must have something to teach us about the search for physical reality. The only interesting thought I know is that maybe evolution by natural selection produced conditioned appreciation of what works for life in the physical realities encountered on Earth, woven into our inheritance from the diverse experiences of many generations of species. And maybe conditioned also is the feeling of beauty of the instruments designed to probe reality (Ivanova and Murphy 2023). 

The failures of assessments from elegance argue against reliance on non-empirical confirmation, but if society remains stable enough there will come a time when theory can no longer be empirically tested and will have to be judged by elegance, logic, and precedents: non-empirical assessments. These assessment might be helped by a better understanding of the connection of physical sciences to our feelings of elegance.

\subsection{The bandwagon effect}\label{sec:bandwagon}

Scientists, being people, are subjects to fads, what Lakatos (1978, p. 91) termed the bandwagon effect. The distinguished astrophysicist Geoffrey Burbidge was fond of the term. The distinguished astronomer Arthur Wolfe, who knew Burbidge well, wrote that (Wolfe 2010)
\begin{quotation}
\noindent 
In my opinion, Geoff's contrarian views
reflected his distrust of the bandwagon phenomenon, which
he interpreted as the uncritical behavior of scientists following
what he regarded as the wrong-headed views of a few elite
leaders.
\end{quotation}
It happens: General relativity was a canonical part of theoretical physics because elite physicists endorsed it. I have heard Geoff's particular complaint that the community acceptance of the general relativity theory of the expansion of the universe was not empirically well supported. He was right prior to the year 2000, when general relativity was more popular than merited by the evidence. But soon after that advances in observations made a persuasive case that this theory is a good approximation to reality (Peebles 2022a, chapter 6; here Sec.~\ref{sec:GR}). 

Andrew Pickering (1984,  p. 7) had thoughts similar to Burbidge's about the bandwagon for QCD:
\begin{quotation}
\noindent by interpreting quarks and so on as real entities the choice of quark models and
gauge theories is made to seem unproblematic: if quarks really are the
fundamental building blocks of the world, why should anyone want
to explore alternative theories?
 \end{quotation}
Pickering was observing the confusion of creation of QCD, the analog of QED for strongly interacting particles, when there still was reason to question the new ideas of quarks and gluons. But QCD has since become a predictive theory that passes the abundant tests reviewed by Campbell, Huston, and Krauss (2018). The bandwagon pointed in productive directions for QCD and relativistic cosmology. 

Bandwagons, like elegance, can head in wrong directions. In the 1990s the community opinion was that the universe is expanding in effect at escape speed, the kinetic energy of expansion exactly balancing the magnitude of the negative gravitational potential energy of the material content of the universe. If so the universe would be predicted to expand into the indefinite future at a rate that is ever slowing by the ever diminishing attraction of gravity. In general relativity it would require that Einstein's cosmological constant, $\Lambda$, vanishes. This has the appeal of a helpful conjecture, that a physical principle to be discovered forces the quantum vacuum energy density to vanish, and with it the value of $\Lambda$. But I, Neta Bahcall at Princeton, and a few others were complaining that the evidence is that the cosmic mean mass density is too small for this picture, a challenge to the bandwagon favoring a negligible value of $\Lambda$. By the end of the century improved tests forced an abrupt change of thinking: the community learned that it had to live with the inelegant $\Lambda$ (from the evidence reviewed in Peebles, Page, and Partridge 2009). 

An aspect of the bandwagon effect, expectation bias, is illustrated by the variation with  time of published values of the measured speed of light. Reports tended to be consistent with other recent previous reports within measurement uncertainties, while the similar values reported at similar times could be inconsistent with later more accurate measurements. Henrion and Fischhoff (1986) show data illustrating this. It is largely an effect of the peer pressure of a sociological bandwagon, the hesitation to present results inconsistent with what already seemed to be known. It is seen also in the history of measurements of Hubble's constant in cosmology, here largely the results of discoveries of systematic errors but also, inevitably, peer pressure.

So what is the lesson? Pay attention to fads; community endorsement can be perceptive. But be wary; the community can be wrong.

\subsection{Sociologies of physical science}\label{sec:SSK}

One camp in the sociology of science draws lessons from observations of what scientists are doing. The sociologist Robert Merton  (1961) discussed an important example in the article {\it Singletons and Multiples in Scientific Discovery: A Chapter in the Sociology of Science}. Merton discussed a common experience in science: when an interesting idea starts to circulate there is a good chance the idea had already been proposed, independently, and escaped wide attention, or else will be independently proposed if news does not travel fast enough.  Merton referred to the sociologists William F. Ogburn and Dorothy S. Thomas (1922), who published a list of 148 examples ``collected from histories of astronomy, mathematics, chemistry, physics, electricity, physiology, biology, psychology and practical mechanical invention.'' 

The phrase, ``the times were right,'' has been dismissed as trite but it fits this phenomenon. Multiples can result from developments of technology that enable scientific or practical advances that more than one person or group might independently recognize and use. But also important is a sociological effect, that half-formed ideas tend to move as gossip through the community by casual remarks and behavior that can be evocative but not recognized until the gossip reaches someone, or maybe more than one person, who is prepared to act on whatever form of the thought came through. Maybe this has something to do with the concept of acausal synchronicity in psychoanalysis. But the simple fact is that people communicate hints of what they are thinking and doing in many ways, by words, gestures, attitudes, and even what is not said. Thoughts float around and occasionally promote action.

Another camp, the sociology of scientific knowledge, or SSK, gives more attention to the inevitable social influences on what scientists are thinking and doing, and in the strongest versions suggest that this social influence causes physics to be a largely social construction. Consider examples of the range of thinking.

Bloor and MacKenzie (1997) expressed a soft version of SSK:
\begin{quotation} 
\noindent the goal of the sociology of knowledge, in our view, is the
explanation of belief, not its evaluation \ldots\ For the historian or sociologist
studying nineteenth-century evolutionism, for example, both Darwinism and anti-Darwinism stand equally in need of explanation.
\end{quotation}
Bloor (1991) added
\begin{quotation}
\noindent But doesn't the strong programme [of SSK] say that knowledge is purely social? Isn't that what the epithet `strong' means? No. The strong programme says that the social component always is present and always constitutive of knowledge. It does not say that it is the {\it only} component, or that it is the component that must necessarily be located as the trigger of any and every change: it can be a background condition.
\end{quotation}
This makes sense. I have mentioned the non-empirical confirmation of general relativity before 1960. I began working on relativistic cosmology in the early 1960s, even though I was dismayed by the scant observational evidence for this subject along with the scant tests of the fundamental theory, general relativity. It helped that I was close to alone in this research; I had little competition. I was encouraged by the tolerant attitudes of colleagues, and the interest in what I was doing by my former thesis adviser, Professor Robert (Bob) Henry Dicke, who I considered my professor of continuing education. This was sociology. But also important was the realization in 1965 that we are in a near uniform sea of microwave radiation with a spectrum that was known to be at least close to thermal. This was something new and interesting to study, and maybe related to cosmology.  

Though Kuhn rejected the strong programme of SSK I place on the more debatable side of this philosophy Kuhn's (1992, p.~7) argument that scientific research seeks
\begin{quotation}
\noindent the facts from which scientific conclusions should be drawn, together with the conclusions---the new laws or theories---which should be based upon them.  These two aspects of the negotiation---the factual and the interpretive---are carried on concurrently, the conclusions shaping the description of facts just as the facts shape the conclusions drawn from them\ldots\ Such a process is clearly circular, and it becomes very difficult to see what role experiment can have in its outcome.
\end{quotation}
This circular reasoning is an important part of the healthy search for hints from experiments and observations that guide considerations of options for revising or devising theories in the search for better fits to data, and it is hoped to reality. More senior physicists usually have greater influence in assessing these considerations: sociology at work. But Kuhn seems to have missed the key point that is illustrated by the experiment mentioned in SSR (pp. 26--27, and discussed here in Section~\ref{existence_tests}) to detect the neutrino. The theory guided the design of the experiment. As Kuhn put it the ``paradigm theory is implicated directly in the design of apparatus.'' This is because the experiment was designed to test the theory. A failure of detection would have been a serious falsification. This is the thinking discussed in Section~\ref{Predictions and falsifications}.

In the book {\it Scientific Knowledge: A Sociological Analysis}, Barnes, Bloor, and Henry (1996) present arguments along a strong SSK line of thought that is to physicists distinctly odd. An example begins with the statement on page 107 of {\it Scientific Knowledge} that
 \begin{quotation}
\noindent the different physical
constants which appear in different problem solutions serve to connect
them together and make the data emerging from the use of one relevant
to the use and appraisal of others. Thus,
the velocity of light [$c$] appears as a universal constant in many exemplars [paradigms, or theories] in physics \ldots\ [$c$ must] have a single value, and this requirement makes the development of the different exemplars in which they appear mutually interdependent and mutually constraining.$^{13}$  
\end{quotation}
The superscript 13 refers to the note on page 207 in {\it Scientific Knowledge} in which the authors ask 
\begin{quotation}
\noindent what is the nature of the need to keep these constants [including $c$]  fixed: what kind of a transgression is it to refuse to keep them fixed? \ldots\  A strictly sociological answer to the second question is that to accept the need to keep certain constants and values fixed is to align one's practice with the practice of other scientists, and nothing more. To share physical constants is to share conventions. To propose alternative values for the constants is to challenge the practice of other scientists.
\end{quotation}
 I trust the authors of {\it Scientific Knowledge} did not realize how insulting their ``strictly sociological answer'' is to the tradition of generations of physicists who spent so much time and effort making precision measurements of $c$, $e$, and the other fundamental constants and carefully compiling and comparing the results with due attention to the inevitable measurement uncertainties. Examples of his great effort are reviewed in Sections~\ref{sec:precisiontests} and~\ref{sec:FundamentalPhysicalConstants}. The authors could have known about this through the celebrated series of papers by Richard Cohen and Jesse DuMond in the late-1940s through to the mid-1960s. If there were a way to ``align one's practice'' to differ from ``the practice of other scientists'' these generations of careful and capable people would have run across  it. It would have been big news.

Barnes et al. (1996) drew evidence for their argument---the scientific tradition allows adjustable results---from the measurements by Robert Andrews Millikan (1917 and earlier references) of the electric charges on drops of water and oil. It drew critics and the ``battle of the electron'' mentioned in Section~\ref{sec:FundamentalPhysicalConstants}. Bad measurements must be rejected, as Millikan did, but it requires great care because rejection of good data or acceptance of bad data could bias results, maybe allowing artificial consistency or inconsistency with other measurements. Experimentalists take great care to control the problem. In double blind medical trials the subjects and the scientists do not know who received the placebo until completion of  the trial. The Higgs particle was detected by two groups that used the same particle accelerator beam line but different technologies applied to different methods of detection and analyses. Both found significant detections with consistent Higgs particle masses, important support for a celebrated result. Peirce (1878) discussed another important check, the comparison of values of fundamental constants derived from measurements of different phenomena by different people, all of whom had to take great care with the rejection of faulty measurements.  Barnes et al. (1996) chose an appropriate topic for sociology; the ``battle of the electron'' is a fair illustration of the way scientists can behave. But the authors did not present a clear example of bad science; compare Birge's (1929) reduction of Millikan's data for the unit of electric charge (eq.~[\ref{eq:Birge_e}]) to the quite similar DuMond and Cohen (1953) result (eq.~[\ref{eq:DuMondCohen}]).

What did the authors of {\it Scientific Knowledge} mean by their ``strictly sociological'' explanation of the consistency of measurements of the fundamental constants?  Surely they did not intend to offer a sociological answer that is manifestly inconsistent with thoroughly explored and established physics. 
I guess they might have presented their ``strictly sociological answer'' as an example of incorrect sociology, but why would they do that, and if they did why would they not explain it? I do not imagine the authors intended to offer the absurd proposal that Cohen and DuMond manipulated the data from the far more capable post-WW~II technology to secure agreement with the values of $c$ and $e$ that Birge compiled in the late 1920s, or the values of $c$ Peirce mentioned in 1878 (Sec.~\ref{Predictions and falsifications}). Many experimental groups contributed results to the Cohen and DuMond compilations. Try to imagine them all keeping what they report consistent with prewar measurements, thus concealing an anomaly that would be a celebrated discovery. 

The authors of {\it Scientific Knowledge} were aware of challenges to Millikan's measurements of the charge of the electron, but they cannot have looked much further into a tradition that is a deeply important part of the science that is the subject of their sociological considerations.

I refer yet again to Thomas Kuhn's thinking because he was trained as a physicist and offered interesting opinions about a subject that he must have known reasonably well. Norton Wise (2016) recalls Kuhn's emphatic disapproval of SSK in the early 1970s. Kuhn (1992, pp. 8,9) later wrote that
\begin{quotation}
\noindent the most extreme form of the movement, ``the strong program,'' has been widely understood as claiming that power and interest are all there are. Nature itself, whatever that may be, has seemed to have no part in the development of beliefs about it.  \ldots\ I am among those who have found the claims of the strong program absurd: an example of deconstruction gone mad.
\end{quotation}
It is easy to imagine why Kuhn objected to this version of SSK; he knew examples of solid progress of physical science, as in the detection of  neutrinos. But Kuhn held to the thought that physicists underestimate the influence of our social conditioning on the lessons we draw from Nature. It is illustrated by what Galison (2016, p. 58) found in Kuhn's notebooks:
\begin{quotation}
\noindent objective observation is, in an important sense, a contradiction in terms. Any particular set of observations \ldots\ presupposes a predisposition toward a conceptual scheme of a corresponding sort: the `facts' of science already contain (in a psychological, not a metaphysical, sense) a portion of the theory from which they will ultimately be deduced.
\end{quotation}
Kuhn was right to argue for the influence of society on the natural sciences, including fundamental physics; it is an inevitable part of how people behave. But the essential thing is the confrontation of theory with the empirical evidence reviewed in Section~\ref{sec:tests} that offers abundant checks that we have quite a useful  approximation to mind-independent reality.

\subsection{Knowledge gained and knowledge lost}\label{sec:empiricalbasis}

Thomas Kuhn argued (in SSR p. 102) that Newton's Laws are not
\begin{quotation}\noindent
 a limiting case of Einstein's. For in the passage to the limit it is not only the forms of the laws that have changed. Simultaneously we have had to alter the fundamental structural elements of which the universe to which they apply is composed \ldots\ The normal-scientific tradition that emerges from a scientific revolution is not only incompatible but often actually incommensurable with that which has gone before.
 \end{quotation}
The philosopher Paul Feyerabend recalled that he and Kuhn independently found the term ``incommensurable,'' taken from mathematics, to be useful in their interpretations of the natures of the natural sciences. (Oberheim and Hoyningen-Huene 2018 review the rich varieties of thinking about {\it The Incommensurability of Scientific Theories}.) Feyerabend and his mentor Karl Popper shared an interest in quantum measurement theory, but I have not found evidence that Feyerabend was exposed to the application of quantum theory to real physical systems that Kuhn experienced. Feyerabend (1970) recalled vigorous debates with Kuhn in 1960-61 when both were at the University of California in Berkeley, along with cases of common thinking. Kuhn (1970b, p. 20) asked why physicists think they have been accumulating ever better approximations to reality, which is our Assumption~C.
\begin{quotation}
\noindent Is it not possible, or perhaps even likely, that contemporary scientists know less of what there is to know about their world than the scientists of the eighteenth century knew of theirs? Scientific theories, it must be remembered, attach to nature only here and there. Are the interstices between those points of attachment perhaps now larger and more numerous than ever before?
\end{quotation}
Feyerabend (1970, p. 219) recalled that Kuhn and he 
\begin{quotation}
\noindent agreed that new theories, while often better and more
detailed than their predecessors were not always rich enough to deal with
{\it all} the problems to which the predecessor had given a definite and precise
answer. The growth of knowledge or, more specifically, the replacement
of one comprehensive theory by another involves losses as well as gains.
\end{quotation}
Kuhn (in SSR p. 6) put it that
\begin{quotation}
\noindent the major turning points in scientific development associated with
the names of Copernicus, Newton, Lavoisier, and Einstein \ldots\ Each of them
necessitated the community's rejection of one time-honored scientific
theory in favor of another incompatible with it.
\end{quotation}

From the physicists' point of view these arguments overlook the fact that empirical knowledge does not go away with the introduction of a new physical theory; the improved theory must take account of the empirical knowledge gained that had been used to test the older theory. Einstein's general theory of relativity would not be interesting if its predictions did not agree with Newtonian gravity when speeds are much less than the speed of light and gravitational potential differences are small. These are the conditions under which the Newtonian theory is thoroughly tested and found to be accurate in the applications to the motions of the planets and their moons, apart from tiny relativistic corrections. Einstein certainly understood this condition; he had the check of consistency with Newtonian physics in his search for the field equation of the general theory of relativity. Useful knowledge was not lost by the replacement of Newtonian gravity with this new theory; it was enriched. For another example consider what happened when Maxwell put aside the search for a mechanical model of the ether and put together two laboratory-based theories, electricity and magnetism. It required  introduction of the hypothetical vacuum displacement current. The lost of the mechanical ether was not regretted because it never produced viable predictions. The two older theories of electricity and of magnetism were not lost; they became useful limiting cases of the unified theory. Another revolution reduced classical electromagnetism to a limiting case of the quantum field theory, QED. It is the subject of Volume 4 in the Landau and Lifshitz {\it Course of Theoretical Physics}. But Maxwell's classical theory is not forgotten; it is reviewed in beautiful detail in the second volume of the Landau and Lifshitz series and still put to productive everyday use. 

An engineer designing a transmission line for electromagnetic energy thinks about electric and magnetic fields with definite values as functions of position and time, an effective ontology for the purpose. A scientist designing an experiment to test QED thinks about the quantum electromagnetic field operator, which might be said to be another effective ontology. But the scientist must also think about the classical magnetic and electric fields that direct the motions of particles, cause detectors to operate, and communicate what is detected to data storage and analysis units and from there to the scientist. The  engineer might not know about QED, and the QED scientist might not be aware of continental drift, regrettable  consequences of the breadth of research in the natural sciences. These are losses to individuals but not to the natural science community. The demotion of the Earth from the center of the world to an ordinary place in a universe that is evolving might be a cultural loss to some, but the Vatican is comfortable with this change of apparent ontology.
 
Science and society have lost useful knowledge. One is handwriting; others range from how to avoid tsunamis to the proper care and use of resources from fields and forests, the ground and the air. But I cannot think of useful knowledge physical science has lost in the past few centuries. 

\subsection{The unity of science}\label{sec:unity}

Thomas Kuhn (1992, pp. 18-20) argued that
\begin{quotation}
\noindent [Natural sciences] should be seen as a complex but unsystematic structure of distinct specialties or species, each responsible for a different domain of phenomena, and each dedicated to changing current beliefs about their domain in ways that increase accuracy and the other standard criteria I've mentioned. For that enterprise, I suggest, the sciences, which must then be viewed as plural, can be seen to retain a very considerable authority \ldots\ what replaces the one big mind-independent world
about which scientists were once said to discover the truth is the
variety of niches within which the practitioners of these various
specialties practice their trade.
\end{quotation}
Kuhn's statement does not make mention of the unity of physics that is so valued by the physics community. Another good challenge is to explain why physicists are so enthusiastic about unification. Physicists feel they have a good answer, but as usual it requires discussion.

Niches in the natural sciences and their practical applications are real and unavoidable. The broad variety of complicated situations requires specialization of knowledge and crafts: niche science. But consider that a farmer with a large investment in the growth of wheat has the benefit of generations of practical experience, but at least as important is the essential help of plant scientists who  know how to breed varieties of wheat that are resistant to pathogens I used to know as ``rust.'' Now plant scientists can genetically engineer wheat for resistance to pathogens and climate changes. If you drive through wheat-growing country in North America you will see signs in wheat fields that identify healthy varieties that seed companies will continue to produce and sell until the inevitable evolution of some pathogen enables it to attack this strain of wheat. Here is an illustration of the fragmentation and unity of science. Farmers need the expertise of plant scientists who need the expertise of molecular biophysicists who study the functions of the large biophysical molecules that help determine genetics, with the physicists' help with models of large molecules based on quantum physics that could be done better. 

Physicists treasure the revolutionary advances in quantum and relativity physics for the additions to the demonstrations of the unity of science, the successful applications of the laws of physics to ever broader ranges of phenomena. Recall in  quantum physics  the demonstrations of quantum entanglement of photons separated by a thousand kilometers; entangled electrons in Cooper pairs that have undefined positions within a superconducting flow a meter around; the properties of atoms and simple molecules; the properties of atomic nuclei inside atoms; the existence of positrons and other antiparticles; and the remarkably successful theory of subatomic physics that has been thoroughly tested down to quarks and gluons. Consider the precision tests of quantum physics reviewed in Section~\ref{sec:precisiontests}, and the great variety of ways to establish tight constraints on values of the fundamental physical constants reviewed in Section~\ref{sec:FundamentalPhysicalConstants}. And bear in mind the many practical applications of this new physics, from light bulbs with great efficiency and durability to cell phones that can help find a missing hearing aid. 

If science is converging to the unique mind-independent reality envisioned in Assumption~D in Section~\ref{sec:startingassumptions} then research in each niche that probes the world in its own way will reach a useful approximation to this unique reality. Apparent inconsistencies among niche theories must be expected because each probe reaches a different approximation to fit what is observed under particular conditions. Sorting this out and checking whether the effects of mind and the limitations of theories and experiments can be taken into account to allow demonstration of convergence of niche theories to a unique reality is a serious challenge and wonderful opportunity for research.

Kuhn in 1992 could have known many of these considerations  when he wrote that the sciences must ``be viewed as plural.'' The thought is correct---research in science requires specialization---but incomplete: missing the unity of physics. Why did Kuhn, an intelligent and thoughtful person, not see the unity that so fascinates scientists? I must leave thoughts about this to those who better understand human behavior. 

\section{How Will Physics End?}\label{sec:final theory}

Assumption~D in Section~\ref{sec:startingassumptions} is that advances of research in fundamental physics are approaching ever better approximations to foundational reality. Popper (1959, p. 452) made two good points about this.
\begin {quote}
\noindent I see no reason to believe that the doctrine of the existence of ultimate explanations is true, and many reasons to believe that it is false. The
more we learn about theories,  or laws of nature, the less do they
remind us of Cartesian self-explanatory truisms or of essentialist definitions. It is not truisms which science unveils. Rather, it is part of the greatness and the beauty of science that we can learn, through our own critical investigations, that the world is utterly different from what we ever imagined---until our imagination was fired by the refutation of our earlier theories. There does not seem any reason to think that this process will come to an end.
\end{quote}
As Popper said, future advances in physics could be incremental, advances that never reach a final theory: turtles all the way down. Or maybe a brilliant conceptual advance will compel acceptance. 

The physics we have now offers two directions for the future; I term them the theory program and the empirical  program. The latter is the route followed in Section~\ref{sec:tests} continuing to broader confrontations of theory and practice. The major effort in the theory program acts on the thought that our present well-tested physical theory might be complete enough to allow physicists to fill in the blanks
well enough to arrive at the ``self-explanatory truism'' that Popper distrusted, but now might be adequately supported by established physics. Let us consider first this theory program. 

Steven Weinberg (1992) wrote that
\begin{quotation}
\noindent I do not mean that the final theory will be deduced from pure mathematics \ldots\ It seems to me that our best hope is to identify the final theory as one that is so rigid that it cannot be warped into some slightly different theory without introducing logical absurdities like infinite energies \ldots\ The final theory may be centuries away and may be totally 
different from anything we can now imagine. But
suppose for a moment that it was just around the corner. What
can we guess about this theory on the basis of what we already
know? \ldots\ The one part of today's physics that seems to me likely to
survive unchanged in a final theory is quantum mechanics. This
is not only because quantum mechanics is the basis of all of our
present understanding of matter and force and has passed extraordinarily stringent experimental tests; more important is
the fact that no one has been able to think of any way to change
quantum mechanics in any way that would preserve its successes without leading to logical absurdities.
\end{quotation}
Arkani-Hamed's (2012) thinking is that
\begin{quotation}
\noindent while we may not have experimental data to tell us about physics near the Planck scale [length scale $\sim 10^{-33}$cm], we do have an ocean of ``theoretical data'' in the wonderful mathematical structures hidden in quantum field theory and string theory. These structures beg for a deeper explanation. The standard formulation of field theory hides these amazing features as a direct consequence of its deference to space-time locality. There must be a new way of thinking about quantum field theories, in which space-time locality is not the star of the show and these remarkable hidden structures are made manifest \ldots by removing spacetime from its primary place in our description of standard physics, we may be in a better position to make the leap to the next theory, where space-time finally ceases to exist.
\end{quotation}
Weinberg and Arkani-Hamed are capable physicists whose views carry weight. 

There is the current problem that, although advances in research are real and significant, they are converging in two directions: elementary particle theory on small scales, relativistic cosmology on large scales. The two peacefully coexist, for the most part, but there is the difference of vacuum energy densities in quantum physics and cosmology, and the challenge of reconciling the unitary evolution of quantum states with the allowed flow of information in general relativity. Polchinski (2022) gives a readable account of the trials and tribulations in this and related aspects of the main theory program of completion of fundamental physics.

The progress of research in M-theory, a successor to string theory that Polchinski indicates might be the ``full quantum theory,'' leads to the thought that the final theory will correctly predict the forms of our established physics but leave the values of the physical constants of nature---the fine structure constant or parameter $\alpha = e^2/{\hbar c}$ and all the rest---to be the values that happen to obtain in our universe in the multiverse. It would then be a task for molecular biologists to explain how the measured values permitted formation of the observed nature of life, and for physicists to explain whether these parameter values would obtain in a habitable universe in the multiverse. And the awkward question would remain: is there a still deeper theory? Might it be found by pursuing other directions of research in fundamental physics? Alternatively, this proposed final theory might be empirically justified by the fact that no comparably viable alternative theory has been found (Dawid, Hartmann, and Sprenger 2015). But how do we judge how thoroughly ``theory space'' has been surveyed for other viable theories? Reflections of this sort are in the book {\it Why Trust a Theory?} (Dardashti, Dawid, and Th\'eault 2019).

The other program, empirical, discussed in Section~\ref{sec:tests}, is the confrontation of what is predicted to what is observed, with attention to the discovery of anomalies that might show the way to better theory. The macroscopic side of fundamental physics is less well tested than the small-scale side, with more approachable questions whose resolution might teach us something of value. What are the physical properties of cosmological dark matter and dark energy, and what are their places in a better standard model of particle physics? The difference of values of $\Lambda$ in quantum and relativistic physics is a clear anomaly. The theory of the angular distribution of the CMB (the thermal cosmic background radiation), baryon oscillations and all, fits the precise measurements, an impressive result, but it requires a value of the Hubble parameter that differs from the directly measured result by 5 to 10 percent. The other parameters required to fit the CMB theory and measurement are consistent with the independent constraints, but that could change as the constraints improve and maybe reveal more anomalies. There also are curious discrepancies with observations in the broadly successful computations of galaxy formation based on the physics and initial conditions of the standard cosmology (Peebles 2022b). These discrepancies might grow into informative challenges. Another question: is canonical physics, Einstein's field equation for general relativity and all the rest, really the same in our solar system and in the immense extrapolation to the scales of length and time of the observable universe? 

An example of this last question is the search for possible changes in the value of the dimensionless fine structure parameter $\alpha$. In standard physics $\alpha$ is constant. The value derived from emission line ratios in spectra of extragalactic objects enables the search for evidence of variation of this nominal constant over cosmic ranges of distance and time (Murphy, Webb, and Flambaum 2007; Jiang, Pan, Aguilar 2024). This approach has not yet yielded confirmed evidence of evolution of $\alpha$. Precision laboratory measurements yield a remarkably tight constraint on the evolution of the local value of $\alpha$ (Chiba 2011). And the laboratory measurements of $\alpha$ continue to be consistent with the value  from applications of basic physics discussed in Section~\ref{sec:FundamentalPhysicalConstants} (Parker, Yu, Zhong, Estey, M{\"u}ller  2018). Since current theoretical ideas about fundamental physics  do not offer the prospect of a definite prediction of the numerical value of $\alpha$, evidence that it is not a fixed number would be big news for the theory program. This is one motivation for continuing these lines of research, along with the natural desire to make the best possible measurements. 

The program of tests of quantum physics by its application to chemistry includes the computations of structures of atoms discussed in Section~\ref{sec:atoms}. Even more demanding computations by methods as close as possible to basic quantum physics, relativistic corrections and all, applied to larger atomic numbers, might prove to be a demanding test of quantum physics. Quantum physics gives a good account of molecular hydrogen, and the fascinating difference of properties of this molecule and hydrogen deuteride, the bound state of a hydrogen atom and its stable heavier isotope. More massive molecules are a lot more complicated. Models such as density functional theory, but refined to the closest feasible approximation to quantum physics, might improve understanding of large biophysical molecules such as the somatostatin that interested Bruno Latour (Sec.~\ref{sec:reality}). The results  would interest molecular biologists. 

The consensus in the physics community is that standard quantum physics will pass the test of application to these complicated systems and more. Recall Weinberg's remark that ``no one has been able to think of any way to change quantum mechanics in any way that would preserve its successes without leading to logical absurdities.'' But we have been surprised before and might be surprised again by empirical evidence that challenges quantum physics.

This empirical program replaces the challenge of assessing the significance of non-empirical assessments in the theory program with the challenge of computation of predictions of standard fundamental physics applied to complicated situations to compare to measurements we might already have. This program likely ends in exhaustion and the usual awkward question, could a greater effort using even more powerful means of computation do even better? But something might turn up.

The end of physics has been announced on occasion. I prefer Peirce's later version of his remark quoted in Section~\ref{Predictions and falsifications}, that
\begin{quotation}
\noindent all the followers of science are animated by a cheerful hope that the processes
of investigation, if only pushed far enough, will give one certain solution to each question to which they apply it.
\end{quotation}
If Peirce's cheerful hope continues to animate research in the natural sciences, including the physics that has so persuasively established empirically real things such as the speed of light, then science has a promising future with many intellectual adventures to come.

\section{Concluding Thoughts}\label{Concluding thoughts}

A theme of this essay is that research into how the world functions on the deepest level we have been able to probe has yielded results that are solid, by and large, and have been applied in technology that has revolutionized everyday life. But let us not forget that these fundamental results are themselves interesting, if to your taste. It is remarkable, and worth knowing, that we can make some sense of the fundamental nature of physical reality. I wrote this essay with interested non-physicists in mind. I hope some will look into it, though they might have to skip over lapses into technicalities, and see the elegance of fundamental physics.  

Paul Feyerabend's (2010, p. viii) opinion---he had many---was that
\begin{quotation}
\noindent science should be taught as one view among many and not as the one
and only road to truth and reality.
\end{quotation}
 But students should have the opportunity to see something of the empirical evidence we have about the fundamental nature of the world. This good science fascinates many non-physicists. I expect Feyerabend meant ``truth and beauty'' to include such things as the pleasure of attending a concert that is to your taste, or taking a walk in the woods. These are important, but they belong in a different essay. And I hope we keep the difference clear to all: the one a matter of taste and choice, the other a matter of painstaking observations and experiments that tightly constrain and establish the theory of what certainly looks like mind-independent reality. The distinction is not absolute---there are many differences of opinion among physicists---but empirical evidence, what is observed, is the essential moderator for disputes in science. I hope the same might be so in society. 

It is good to question authority in physics but it can go too far. I admire Ernst Mach's expositions that demonstrate the broad predictive power of classical physics, but cannot understand his lack of appreciation of predictions that he understood so well and are such elegant demonstrations of the unity of physics. A theory that produces successful predictions cannot be all bad (with thanks to W. C. Fields). I admire Thomas Kuhn's independence of mind in exploring the undoubtedly real influence of cultural norms on the development of physics, but I am perplexed by his lack of appreciation of the unity of physics exemplified by the broad range of its successful predictions. Why Mach and Kuhn were skeptical about established opinions in the physics community is a puzzle to physicists, and maybe a topic for sociologists to examine from the skeptic and establishment sides. 

Physicists are quite capable of fixed opinions that can be taken up by enough to be termed the bandwagons discussed in Section~\ref{sec:bandwagon}. An example is the opinion in the 1990s that Einstein's cosmological constant $\Lambda$ surely is negligibly small. One reason was (and still is) that there is no natural place for it in our fundamental theory. Where is the unification? The anthropic principle (Sec.~\ref{sec:anthropic}) offers a way out, but a more elegant theory might be found and the question then might be why we should trust a theory that cannot be tested. The question is heard about other theories, and will be increasingly common. But that will not be the end of empiricism in fundamental research. It will continue in the essentially unlimited task of extending the applications of fundamental physics to complicated situations in the encroachment of fundamental physics into the other natural sciences, always to be compared to what is observed.

Finally, let us bear in mind that fundamental physics is exploring a tiny slice of human knowledge and experience, and a dry slice at that. The approach has been seriously productive, but where are the floods of words needed to describe the subtle shades of meanings of the world's immense varieties of phenomena, from the simpler chemical procedures to the self-awareness of spectacularly complex human beings? All this is present in the many layers of reality that rest on what physicists think they are empirically establishing but cannot prove or use to explain all the wonderful complexity of the world. Explorations of such things will continue, and research in the physics described in this essay will continue to fascinate for its concrete reasons to think that our world rests on a fundamental physical basis that we can explore and come to understand, in approximations. 

 \section{Acknowledgements}
 
I am grateful to David Hogg for the many discussions that led me to write this essay. Discussions with Tony Rothman and his careful readings of drafts of this essay helped me clarify what I thought I meant to say and forced me to say it more clearly in a more sensible order. Victor Albert, George Efstathiou, Will Happer, Peter Lindenfeld, David Mermin, and Timur Tscherbul answered my questions and helped  straightened my thinking. The appearance of an earlier version of this essay on arXiv yielded useful advice from Bryce Cyr, Jorge Ernesto Horvath, Peter Morgan, Boud Roukema, Angelo Vulpiani, and James Wells. 

I am not complaining that I have not received financial support for this essay, and I am grateful to  Princeton University for its valuable support by providing me an office.

\label{lastpage}
\end{document}